\documentstyle[aas2pp4]{article}       

\lefthead{Cavaliere, Menci \& Tozzi}
\righthead {Groups and Clusters of Galaxies}

\begin{document}

\title{Diffuse Baryons in Groups and Clusters of Galaxies}

\author{A. Cavaliere}
\affil{Astrofisica, Dip. Fisica, II Universit\`a di Roma,\\
via Ricerca Scientifica 1, 00133 Roma, Italy}

\author{N. Menci}
\affil{Osservatorio Astronomico di Roma,
via Osservatorio, 00040 Monteporzio, Italy}

\and

\author{P. Tozzi}
\affil{ Astrofisica, Dip. Fisica, II Universit\`a di Roma,\\
via Ricerca Scientifica 1, 00133 Roma, Italy}

\begin{abstract}

To predict the X-ray observables associated with the diffuse baryons
in clusters of galaxies, we develop a new physical model for 
such a hot intra-cluster plasma.  Our framework is provided by
the hierarchical clustering cosmogony for the dark matter, and by 
the standard FRW or Lema\^itre cosmologies constrained by cosmic ages.  

As to the plasma dynamics and thermodynamics we propose a 
semi-analytical approach based on {\it punctuated equilibria}.  
This comprises the following blocks that we compute in detail: 
Monte Carlo ``merging histories'' to describe the dynamics of 
dark matter condensations on scales of order $1-10$ Mpc, and the 
associated evolution of the gravitational potential wells; 
the central {\sl hydrostatic} 
disposition for the ICP, reset to a new equilibrium after each merging 
episode;  conditions of shock, or of closely adiabatic compression at the 
{\sl boundary} with the external gas, preheated by stellar energy feedbacks. 
Shocks of substantial strength are shown to prevail at the outskirts
of rich clusters in a universe with decelerated expansion.  

From our model 
we predict the $L-T$ relation, consistent with the data as for shape and 
scatter.  This we combine with the mass distribution provided 
by the canonical hierarchical clustering; the initial perturbation spectra 
are dominated by Cold Dark Matter but include enough baryons
to account for the high abundance sampled by the X--ray clusters, 
and are COBE--normalized.  
Thus we predict the $z$-resolved luminosity functions, with 
the associated source counts and redshift distributions. 
We predict also the complementary contribution by 
the unresolved groups and clusters to the soft X-ray background. 

These results are compared with two recent surveys from ROSAT; one defines
the local luminosity function over nearly three decades of $L$, and the other
shows little or no evolution out to $z\sim 0.8$.

Our results confirm that the critical cosmology coupled 
with Standard CDM is ruled out by its overproduction 
of local clusters. On account of underproduction, instead, 
we rule out open cosmologies (the cheapest way to solve the baryonic 
crisis and to freeze evolution), 
except for a narrow range around $\Omega_o =0.5$; even there, we find the 
consistency with the full data base to be hardly marginal.  For the  CDM 
cosmogony with $\Omega_o=0.3$ but in flat geometry, 
we obtain acceptable fits.  For the tilted CDM 
perturbation spectrum with high baryonic content in the critical universe, 
we obtain marginal consistency.  
The cosmogonical/cosmological sectors of the cluster history 
 are independently 
testable by means of a lower bound to the evolved temperature distribution, 
as can be measured with SAX and XMM out to moderate $z$. 

Finally, we discuss the effective limitations of X-ray clusters and groups 
as cosmological signposts, and their brighter prospects toward 
the astrophysics of the ICP and the cosmogony of large, high--contrast 
structures. 

\end{abstract}


\section{Introduction}

Groups and clusters of galaxies constitute cosmic structures 
sufficiently close to equilibrium and with 
sufficient density contrast ($\delta\approx 2\, 10^2$ inside the virial radius 
$R$) as to yield definite 
observables, and possibly to offer reliable signposts for cosmology. 

They stand out to substantial depths of space-time 
not only in the optical band, but even more in X--rays. This 
is because their gravitational potential wells, shaped by a dominant dark mass 
$M$, contain not only baryons condensed into stars but also a 
larger amount of {\it diffuse baryons}. The latter, with densities 
$n\sim 10^{-3}$ cm$^{-3}$ and virial temperatures 
$k\,T\sim GM\,m_H/10 R\sim 5$ keV in rich clusters, satisfy the plasma 
condition $kT/ e^2\,n^{1/3}\gg1$ exceedingly well; in fact, they
do by a factor $10^{11}$, vastly larger than for the baryons 
inside the stars. Such a hot {\it intra--cluster plasma} (ICP) emits powerful 
X-ray luminosities $L\propto n^2\,R_X^3\,T^{1/2}\sim 10^{44}$ erg/s
by optically thin thermal bremsstrahlung from central regions of overall 
 radius $R_X\sim 1$ Mpc.  

The temperature directly probes the height of the 
potential well, with the baryons in the  role of  mere tracers; 
on the other hand, the luminosity 
with its strong dependence on density  reliably 
probes the baryonic content. Statistically, 
a definite  $L-T$ correlation is observed (albeit with considerable scatter), 
and this provides the crucial 
link to relate the X-ray luminosity functions with 
the statistics of the dark mass $M$ or with that of the corresponding $T$. 

But groups and clusters are intrinsically complex systems. To begin with, their 
dynamical history is marked by extensive, repeated {\it merging} of clumps, 
both in the form of nearly isotropic accretion of small units, 
and in the form of a few large, anisotropic coalescence events.  
This is shown in detail by all N-body simulations (see, e.g., 
Schindler \& Mueller 1993; Tormen 1997; 
Roettiger, Stone \& Mushotzky 1997), and is increasingly indicated by the 
data (see, e.g., Zabludoff \& Zaritsky 1995; Henriksen \& Markevitch 1996; 
Jones et al. 1997).  
 
The timing of such a dynamical evolution (and specifically the present 
merging rate) is set by the {\sl cosmological} framework.  
Adopting the homogeneous isotropic 
FRW or Lema\^itre cosmologies, the expansion is 
parametrized by the Hubble constant $h$ (in units of 100 km/s Mpc), and by the 
density parameter $\Omega_o$ (to which the baryons contribute the  
fraction $\Omega_B$); a  possible additional 
contribution is given by $\Omega_{\lambda}$ associated 
to the cosmological constant. 

Given the cosmological framework, the {\sl cosmogony} (i.e., 
the process of structure formation) is treated in terms of the standard 
hierarchical clustering scenario (see Peebles 1993),  
where all structures form by gravitational instability of initial density 
perturbations $\delta$ in the dark matter (hereafter DM).  
Important parameters are the shape of their power spectrum 
$\langle |\delta_k|^2\rangle\propto k^{n_p}\,T^2(k)$ at the recombination,  
and the normalization measured at large scales by COBE/DMR (Gorsky et al. 
1996). 
The transmission function $T(k)$ depends on the specific model assumed for the
DM; given this, 
the amplitude $\sigma_8$ (i.e., the normalization extrapolated down to 
the relevant scale $8\,h^{-1}$ Mpc) and 
 the effective spectral slope $n_e$ 
depend on $\Omega_o$ and $h$. These parameters also affect 
the growth factor $D(z)$ of the perturbations, 
which enters the actual predictions for 
the mass distributions of the condensed clusters or groups (see Appendix A). 

Further complexity is added by the {\sl physics} of the ICP, and this 
constitutes our main aim here.  To now, the ratio of the ICP to DM, and 
specifically its central density $n_o$ and its effective radius are 
poorly understood. 
But the former is especially important, as the factor $n_o^2$ 
enters the luminosities and {\it amplifies} the observed variance. 

Observational information on 
the ICP and on the underlying DM dynamics is provided by the 
local $L-T$ correlation. At higher $z$, further statistical 
information is provided by 
the evolution of the X-ray luminosity function 
$N(L,z)$, or by its integrals like the source counts or the 
$z$-distributions. The problem is that the predictions of observables involve 
not only the ICP physics and the cosmogony (with their intrinsic variances) 
but also cosmology again (with its uncertainties), and so 
the various aspects are not easily disentangled.

The sharpest result obtained so far rules out 
the attractively simple assumption by Kaiser (1986) that the ICP amount 
be proportional to the DM's from groups to clusters at all $z$ and $M$. 
A large number of subsequent works 
(of which we cite here the recent Mathiesen \& 
Evrard 1997, Kitayama \& Suto 1997, and Borgani et al. 1998) 
dealt with the combinations of 
these three {\it sectors}, namely,  cosmology, cosmogony and ICP 
physics.  But  most of these papers,
 following the start by Cavaliere \& Colafrancesco (1988), 
approached the problem by parameterizing the  dependences of the ICP/DM ratio 
on $M$ and $z$, e.g., in the form $L\propto M^{p}\,(1+z)^{s}$.  
While the parameters $p$ and $s$ are constrained to some extent 
by the local $L-T$ correlation, nevertheless there still remains  a
substantial {\it degeneracy} (Oukbir, Bartlett \& Blanchard 1997) between 
ICP physics and cosmology/cosmogony. That is to say,  different 
combinations of $h$, $\Omega_o$, $\Omega_{\lambda}$, $\Omega_B$, $\sigma_8$, 
$n_p$, $s$, $p$ provide close fits to the observables. Conversely, when the 
ICP parameters are varied, different cosmologies appear to be 
preferred by the data; the trade off particularly concerns $\Omega_o$ and 
$ s$, 
which directly govern the global and the ICP evolution, respectively.

To go beyond such degeneracies a 
{\it physical} model is needed for the ICP, including the above 
complexities. Such a 
model must include: the histories of DM halos with their 
hierarchical merging events; the infall of the gas with the ensuing 
compression and shocks; the disposition of the ICP in the potential wells; 
 its conditions at the boundary with the surrounding environment, 
which is modulated by the large scale structures and by stellar preheating. 
A model accounting for all the above is missing so far.  
We stress that the simulations using advanced hydrodynamics coupled with 
N--body codes hardly reach at present enough dynamic range 
(as discussed by Bryan \& Norman 1997) to 
describe DM and ICP over the full range from $\sim 50$ Mpc associated with 
the large scale structures (which guide the ongoing mergers of DM halos), 
to the inner 50 kpc where the ICP yields a considerable contribution to $L$. 
On the other hand, such  high-resolution simulations do not include yet 
the stellar preheating with its crucial effects on the LT relation. 

This motivates us to 
develop here a semi-analytical model which includes, though in a 
simplified form, the features listed above. 
We describe the cluster evolution as a sequence of {\it punctuated equilibria} 
(PE); that is to say, a sequence of hierarchical merging 
episodes of the DM halos (computed with Monte Carlo simulations), associated 
in the ICP  to shocks of various strengths 
(depending on the mass ratio of the merging clumps), 
which provide the boundary conditions for the ICP to 
 re-adjust to a new hydrostatic equilibrium. 
We show that our PE model predicts density and temperature profiles and 
the $L-T$ relation for clusters and groups consistent with the recent data. 

We then use our PE to  predict the 
counts of resolved sources $N(>F)$ for faint fluxes down to $F >  2\,10^{-14}$ 
erg/s cm$^2$, now accurately measured by Rosati et al. (1997). 
We predict also the {\it complementary} observable constituted by the 
contribution of the unresolved groups and clusters to the soft 
X-ray background (XRB). We predict the $z$-distributions, 
and finally 
the full $z$-resolved luminosity 
functions to be compared with recent and  
forthcoming surveys.  

The paper is organized as follows.  In \S 2 we present and discuss our 
approach to the ICP astrophysics.  In \S 3 we give the X-ray observables in 
the form suited to our hierarchical clustering computations; details
 are supplied in the Appendixes A, B and C.  In \S 4 we present the 
results from our approach. The final 
\S 5 is devoted to discussions and conclusions.  

\section {The punctuated equilibria for the ICP}

The X-ray luminosity of a cluster with radial 
temperature profile $T(r)$ and density 
profile $n(r)$ is given by 
\begin{equation}
L\propto \int d^3 r\,n^2(r)\, T^{1/2}(r)~;
\end{equation}
the integration is over the emitting volume $r^3\leq R^3$.  
Expressing $R$ from $T\propto M/R\propto \rho\,R^2$, 
eq. (1) recast into the form 
\begin{equation}
L \propto \overline{[n(r)/n_1]^2}\,T^2\,\rho^{1/2}~, 
\end{equation}
where $\rho$ is the internal DM density, 
$n_1$ is the density just exterior to the cluster boundary, and 
the bar indicates the integration over the cluster volume normalized to 
$R^3$. Note that eq. (2) applies in the isothermal case; the 
corresponding expression for a polytropic ICP is given in Appendix B.

The simplest approach to the ICP state is that adopted in the self--similar 
model (Kaiser 1986) where $n \propto\rho$ is assumed, independently 
of $T$; then $L\propto T^2$ obtains from eq. (2).  
The result conflicts with 
the observed correlation for rich clusters 
 which is close to $L\propto T^3$ (Edge \& Stewart 1991; 
 Mushotzky 1994; Tsuru et al. 1996; Mushotzky \& Scharf 1997).  
Also, when combined with the standard hierarchical cosmogony, the assumption
yields unacceptable fits to the local luminosity function (see, e.g., Kitayama
\& Suto 1997).  Finally, it would predict the 
clusters at higher $z$ to comprise not only 
denser DM, but also equally denser ICP; since bremsstrahlung depends on $n^2$, 
this would imply a strong positive evolution of $N(L,z)$, which is certainly 
{\it not} observed. Rather, the analysis of the ROSAT Brightest Cluster Sample 
(Ebeling et al. 1997) and of the ROSAT Deep Cluster Survey 
by Rosati et al. (1997), extending and strengthening the data by 
Collins  et al. (1997) and by  Nichol et al. (1997), 
indicate no evolution out to $z\approx 0.8$ for $L<3\,10^{44}$ erg/s;
earlier surveys (EMSS, Henry et al. 1992) suggested even a 
(marginal) negative evolution of the bright clusters. 

So to derive the true {\it scaling} of $L$ with $T$ we need a closer analysis 
of the ICP disposition relative to the DM.  
We propose a new approach based upon two 
cornerstones: the profiles $n(r)$ and $T(r)$ are given by the {\it hydrostatic} 
equilibrium with the gradually changing gravitational potential; 
their normalizations, and so the central density, are set by the conditions at 
the {\it boundary} with the external medium. 

The equilibrium profile may be effectively represented 
with a polytropic relation starting from the cluster 
boundary at $r_2$, say: 
\begin{equation}
n(r)/n_1=g(T)\,\Big[{T(r)\over T_2}\Big]^{1/(\gamma-1)}~.
\end{equation}
Here $T_2$ is the temperature just interior to the boundary; we conveniently 
 use $g(T)\equiv n_2/n_1$ for the ratio of the interior to the exterior 
density, 
to include any shock discontinuity at the boundary. The appropriate 
values for $\gamma$ will be discussed in \S 2.1 and \S 2.4. 

Actually, the ICP is reset to a {\it new} equilibrium after each episode of 
accretion or merging of further mass.  In our PE approach,  
the history of such episodes is followed in the framework of the 
hierarchical clustering by Monte Carlo simulations, as explained below. 

Two relevant {\it limiting} forms of eq. (3) are constituted by 
by the ``shock'' model of CMT97, and by the ``adiabatic'' models of Kaiser 
(1991) and of  Evrard \& Henry (1991). In both the outer gas 
is expected to be 
{\sl preheated} at $T_1\lesssim 1 $ keV 
(Ciotti et al. 1991; David et al. 1993, 1995; Renzini 1997) 
by $z\lesssim 2$, due to 
feedback energy inputs following star formation and evolution. 
Preheating temperatures 
$T_1\gtrsim 0.1$ keV also would prevent the cooling catastrophe 
from occurring, see White \& Rees (1978); 
Blanchard, Valls Gabaud \& Mamon (1992). In point of fact, Henriksen \& 
White (1996) find from X-rays evidence for diffuse cool gas at $0.5 - 1$ keV 
 in the outer regions of a number of clusters.  In the present context, 
preheating will inhibit the attainment 
 of the universal baryonic density in gravitational wells with virial 
temperatures comparable to $T_1$.

These limiting models differ in their treatment of the 
boundary conditions and of the merging histories.

\subsection{Shocks and hydrostatic equilibrium}

The key boundary condition is provided by the dynamic stress balance 
$P_2=P_1+m_H\,n_1\,v_1^2$, relating the exterior and interior 
pressures $P_2$ and $P_1$ to the inflow 
velocity $v_1$ driven by the gravitational potential at the boundary. 
We expect the inflowing gas to become 
supersonic in the vicinity of $R$, when $m_H\,v_1^2> 2 kT_1$.  
In fact, many hydrodynamical simulations of loose gas accretion 
into a cluster (from Perrenod 1980 to Takizawa \& Mineshige 1997) show 
shocks to form, to convert most of the bulk energy into thermal energy, 
and to expand slowly 
remaining close to the virial radius for some dynamical times.  So we 
take $r_2\approx R$ (which follows the structure growth, since 
$R\propto M^{1/3}$), and focus on nearly static conditions inside, with 
$v_2^2<< v_1^2 $.  

The post-shock state is set by conservations across the shock 
not only of the stresses, but also of mass and energy, as described by the  
Rankine-Hugoniot conditions (see Appendix B). 
These provide at the boundary the temperature jump $T_2/T_1$, and the
corresponding density jump $g$ which  reads 
\begin{equation}
g\Big({T_2\over T_1}\Big) = 
2\,\Big(1-{T_1\over T_2}\Big)+\Big[4\, 
\Big(1-{T_1\over T_2}\Big)^2 + {T_1\over T_2}\Big]^{1/2}
\end{equation}
for a plasma with three degrees of freedom.
Eq. (4) includes both {\it weak} and {\it strong} shocks.  For weak shocks 
with $T_2 \approx T_1$ (appropriate for small groups accreting 
preheated gas, or for rich clusters accreting comparable clumps), this 
converges to the truly adiabatic 
relationship $n_2/n_1= (T_2/T_1)^{3/2}$ up to second order inclusive, see 
Landau \& Lifshitz (1959).  On the other hand, it is shown 
in Appendix B that for strong shocks (appropriate to "cold inflow" as in rich 
clusters accreting small clumps and diffuse gas) the approximation 
$k\,T_2\approx -\phi_2/3+3k\,T_1/2$ holds, where $\phi_2$ is the gravitational 
potential energy at $r_2\simeq R$.  

Inside $R$, the temperature and density profiles $T(r)$ and $n(r)$ are 
matched to $T_2$ and to $n_2$ by {\it polytropic} profiles or by their
{\it isothermal}  limit.  We numerically compute such profiles using the 
hydrostatic support of pressure against gravity (see Appendix B);  for 
definiteness, we use the 
Navarro et al. (1996) representation for the potential and for the 
velocity dispersion
 (which varies slowly in the relevant region).  

Let us consider first for reference 
the simple analytical approximation provided by the standard 
isothermal model 
\begin{equation}
n(r)/n_2=[\rho (r)/\rho_2]^\beta \, ,
\end{equation}
(Cavaliere \& Fusco-Femiano 1976) with the canonical exponent
$\beta \equiv \mu m_H \sigma/kT_2$; here $\sigma$ is 
the one-dimensional velocity dispersion at $R$, and $\mu$ is the 
average molecular weight. 
For the purpose of the analytical approximation we may use 
the strong shock limit for $T_2$, and find 
numerical values for $\beta$ ranging from about $ 0.5$
for groups to 0.9 for rich clusters (a trend consistent
with data being collected by M. Girardi and coauthors, private 
communication).  This implies 
that $R_X/R$ is larger in the former than in the latter (see CMT97).  

The full numerical computations using the expression for $T_2$ given in 
Appendix B, and $\sigma (r)$ and $\phi (r)$ from Navarro et 
al. (1996) confirm this trend and give results shown in fig. 1.  
Note that even in the isothermal case the 
emission-weighted temperature (integrated along the line of sight) 
declines outwards but only very slowly.  

The observed  stronger decline requires a  polytropic equilibrium, where 
the run of $T(r)$ steepens with the index $\gamma$ increasing from 1. 
 In the Appendix B we recall the basic relations, 
and show that the variations induced 
in the volume-averaged luminosity by  increasing $\gamma$ are small. 
Thereafter, we adopt the value $\gamma=1.2$, which also yields for rich 
clusters an integrated baryonic fraction  0.15 out to $R$. 
The result for the emission-weighted $T(r)$ (see fig. 1) is 
a mild decrease out to $r\sim$ 1 Mpc in agreement  with the observations 
(Hughes, Gorenstein \& Fabricant 1988;  Honda et al. 1997; 
Markevitch et al. 1997),  followed by a sharper drop as indicated 
by state-of-the-art 
simulations (e.g., Bryan \& Norman 1997). Note that 
in fig. 1 the discontinuity at the shock has been smeared out 
to a smooth drop by 
the finite resolution, taken at 100 kpc for comparison 
with the simulations and with the forthcoming observations.

\subsection{Merging histories and the $L$-$T$ correlation}

The luminosity $L\propto g^2~ \overline{n^2(r)/n_1^2}~ T^2 \; \rho^{1/2}$
is statistically 
affected by the {\it merging histories} as follows.  For a cluster or group 
of a given mass (or temperature), the effective compression factor squared 
$\langle g^2 \rangle$ is obtained upon averaging eq. (4) 
over the sequence of the DM 
merging events; in such events, $T_2$ is the virial temperature of 
the receiving structure, and $T_1$ is the higher  between the stellar 
preheating temperature and that from ``gravitational''
 heating, i.e., the virial 
value prevailing in the clumps being accreted. 

All that is accounted for in our model 
using Monte Carlo simulations of the hierarchical growth of the DM halos; these 
 are based on merging trees corresponding to the excursion set approach of Bond 
et al. (1991), consistent with the Press \& Schechter (1974) statistics 
(see CMT97).  The averaging procedure is dominated by the events occurring 
 within the last few dynamical times; it results in 
lowering $\langle g^2 \rangle$ 
compared to $g^2$, because in many events the
accreted gas is at a temperature higher than the minimum preheating
value $T_{1\ell}\approx 0.5 $ keV.  In addition, an intrinsic {\it variance} 
is generated, reflecting and amplifying 
  the variance intrinsic to the merging histories.  

The net result is shown in fig 2, and commented in its caption. In agreement 
with the observations (Edge \& Stewart 1991; Mushotzky 1994;  
Tsuru et al. 1996; Ponman et al. 1996),  
the {\it shape} of the average $L-T$ relation 
flattens from $L\propto T^5$ at the group scale (where the nuclear 
energy from stellar preheating 
competes with the gravitational energy from infall) to $L\propto T^3$ at the 
 rich cluster scales. At higher temperatures the shape asymptotes to 
$L\propto T^2$, the self-similar scaling of pure gravity. 
Notice the intrinsic {\it scatter} 
due to the variance in the dynamical merging histories, but amplified by the
$n^2$ dependence of $L$. 

The average normalization formally rises like $\rho^{1/2}(z)$, where $\rho$
is the effective external mass density which increases as $(1+z)^2$ 
(Cavaliere \& Menci 1997) in filamentary large scale structures hosting most 
groups and clusters (see Ramella, Geller \& Huchra 1992).  
This implies a factor 1.3 at $z=0.3$, consistent with the 
observations by Mushotzky \& Scharf (1997).  
Further weakening of the $z$-dependence will 
comes from the increasing depth of 
the central $\phi_o$ for distant 
structures of given $M$, as predicted by Navarro, Frenk \& White (1996).

\subsection {The adiabatic models}

At the other extreme, the models by Kaiser (1991), 
and Evrard \& Henry (1991) 
obtain from PE under two limits, appropriate only for rapidly expanding 
universes, as we discuss below. The first limit correspond to 
no currently active merging, with shocks moving outward and vanishing. 
In such conditions, at the boundary  $T_2 \approx T_1$ holds; 
with $T_1$ staying nearly constant after the dynamical freeze 
out, this implies $g \equiv n_2 /n_1 \approx 1$. 
Correspondingly, the central density scales approximately as 
$n_o\propto n_1\,(T_o/T_1)^{1/(\gamma-1)}$.  

The ``adiabatic'' models require also a second limit, concerning the 
internal gas distribution. 
The value $\gamma = 5/3$ is taken at the center, but an isothermal 
$\beta$ profile is assumed (with a fixed $\beta$), 
based on a King--like DM distribution. 
A constant baryonic fraction at $R$ is then required, and this forces 
the core radius to scale as $r_c \propto M^{1/3-1/3\beta}$. 
Thus $L\propto T^{2+3(2-1/\beta)/2}$ obtains, with the
normalization $\rho^{1/\beta -3/2}(z)$. 

Finally, the value of $\beta$ is chosen as an input.  The choice $\beta =2/3$ 
for both clusters and groups leads to the model of Evrard \& Henry (1991), in 
which $L \propto T^{11/4}$ obtains, with constant normalization.  
The choice $\beta 
=1$ leads to the somewhat different model of Kaiser (1991), in which $L\propto 
T^{3.5}$ obtains, with the normalization anti-evolving like 
$(1+z)^{-3/2}$; this will become even more negative when the evolution of 
$\phi$ is taken into account, so conflicting with the data 
of Mushotzky \& Scharf (1997).  

\subsection{ICP models and cosmology}

Shock and adiabatic models can be characterized in terms of entropy
(see Bower 1997).  
Actually, the modes of entropy production and distribution 
correlate with the global dynamics.  

Collapses, merging and the induced shocks are currently 
ongoing in the {\it critical} universe, so that strong shocks form 
close to the virial radius.  
Entropy is continuously  generated in the outer regions, so that its 
radial distribution is raised outwards. Then 
the effective $\gamma$ will be close to one, leading to a roughly flat 
$T(r)$ inside the shock.  The density is determined by the boundary 
conditions after eqs. (4).  Shocks are weaker in groups, the density 
profile is shallower, and the $L-T$ relation steeper.

Conversely, in an {\it open} universe most dynamical action is moved back 
to early times: 
merging and mixing occurred early on, and then subsided; shocks had time to 
expand beyond $R$ and weaken; 
correspondingly, the accretion petered out under nearly adiabatic 
conditions for groups  and for clusters as well. 
Then the effective $\gamma$ is closer to $5/3$, and this may be used 
to roughly scale the central densities with different virial
temperatures $T$  to obtain $L\propto T^{3.5}\,R_X^3$.  The two adiabatic
models adopt additional, and different, assumptions concerning 
$r_c$ or $R_X$, i.e., $r_c\propto 
T^{-0.25}$, or $R_X=const$, as discussed in \S 2.3.  

For open cosmologies with $\Omega_0\approx 0.5$, or for flat ones with 
$\Omega_{\lambda}=1-\Omega_0$, the present deceleration is {\sl intermediate} 
between the two above cases, and the applicability of the 
shock or of the adiabatic model is not so clearcut; 
we shall consider both, finding similar results as is expected. 

\section{Statistics of condensations, and X-ray observables}

We use the dark mass $m\equiv M/M_{o}$ normalized to the characteristic 
mass $M_{o}=0.6\, 10^{15}\,\Omega_0\,h^{-1}~M_{\odot}$ defined in the 
hierarchical clustering theory, see Appendix A and the analytical details 
given in Appendix C.  The X-ray temperature $T$ reads: 
\begin{equation}
T=T_{o}\,m^{2/3}\,(1+z)~.
\end{equation}
On the basis of \S 2, for the shock model the bolometric luminosity is given by 
\begin{equation}
L= L_{o}\,\langle g^2(T)\rangle \, 
\overline{n^2(r)/n^2_2}\,(T/T_{o})^2\,(1+z)~;
\end{equation}
the ratio $\overline{n^2(r)/n^2_2}$ (integrated over the cluster
volume) and the 
factor $\langle g^2 \rangle$ (averaged over the merging histories) have been 
 derived in \S 2.2.  

The constant $k\,T_{o}$ takes on the value $4.5\,\Omega_0$ keV (see 
Appendix C).  The 
luminosity $L_o$ is calibrated on the height of 
the observed, local $L-T$ correlation (see fig. 2)
rather than computed a priori, in view of the subtleties discussed in \S 5.  
At 4.5 keV we find $L_{o} = 1.6\,(1\pm 0.25)\; h^{-2}\; 10^{44}$ erg/s. 
The height of the local luminosity function provides an independent value for 
the normalization, 
which we find consistent with the former to within $15$\%.  

The statistics at different $z$ of DM halos in the HC theory is 
provided by the standard Press \& 
Schechter (1974) formula, which in comoving form ($\rho_o$ being the local 
 cosmological density) reads:  
\begin{equation}
N(m,z)=\sqrt{{2\over \pi}}\;{\delta_{c}\,\rho_o\over \,M^2_{o} }\;
|{{d\ln \sigma}\over {d\ln m}}|\; {m^{-2}\over {\sigma(m)\,D(z)}}\;
e^{-{\delta_{c}^2\over {2\,\sigma^2(m)D^2(z)}}}~,
\end{equation}
where $\delta_{c}$ is the critical threshold for the 
collapses of the density perturbations (depending weakly 
on the cosmological parameters); on the other hand, $D(z)$ is the linear
growth factor for the density perturbations, sensitively depending on the
cosmological parameters (see Appendix A). 
The linear, time-evolved mass variance  $\sigma(m)\,D(z)$ is 
usually represented in the form $\sigma_8 m^{-a}D(z)$ 
where $a\equiv (n_{e}+3)/6$ contains the effective slope 
$n_{e}(M)\approx -1.3\div -2$ 
of the power spectrum $\langle |\delta_k|^2\rangle$ at scales 
$\sim 10 \div 1 $ Mpc. 

It is characteristic of eq. (9) to comprise two kinds of evolution:
the number increase $\propto D^{-1}(z)$ 
at the low--$M$ end, and the shift toward smaller $M$ 
of the upper exponential cutoff.  Such dynamical evolutions, 
modulated by 
cosmology, must combine with the ICP evolution to yield closely constant 
luminosity functions
as observed.  We conservatively adopt 
the Press \& Schechter rendition of the hierarchical clustering, keeping in
mind its problems and limitations (see Bond et al. 1991, Cavaliere et al. 
1993) and deferring discussions to \S 4.4 and 5. 

From such statistics of dark halos using eq. (8) we compute the luminosity 
function $N(L)=N(M)dM/dL$, the expected flux counts $N(>F)$ of X-ray clusters, 
and their contribution to the soft XRB.  The latter two observables read (see 
Appendix C)  
\begin{equation}
N(>F)=R_H\int_{m_1}^{\infty} dm\int^0_{z_F} 
d\omega\,d\ell (z)\,N(m,z)\,dm~,
\end{equation}
$$
I(\nu_o)  =  \int\int_{m_o}^{m_1}R_H\, d \ell(z) N(m)\,dm
\,\overline{[n(r)/n_1]^2}$$
\begin{equation}
~~~~{L_{o}m^{4/3}(1+z)^{5/2}\over 
4\pi\,\Gamma (0.6)\, k\,T_{o}\,m^{2/3}  }\,
\Big[{h\nu \over k\,T_{o}\,m^{2/3}}\Big]^{-0.4}\,
e^{-{h\nu\over kT_{o}\,m^{2/3}}}~.
\end{equation}
Here the lower limit $m_o$ 
is set by the requirement that $T(m,z)>0.5$ keV, the effective lower bound 
 for group temperatures from preheating; smaller masses would 
correspond to galaxies, where the amount of diffuse baryons and the emission 
(per unit total mass) drop sharply (Fabbiano 1996). 

The limit $m_1$ in eq. (10) is set by the maximum between $T(m,z)>0.5 $ keV 
and the limiting flux of cluster or group sources, 
for which we conservatively adopt $4\,10^{-14}$ erg/s cm$^2$. 
On the other hand, $m_1$ also constitutes the upper limit for the 
 the {\it unresolved} sources contributing to the XRB.  

Thus a {\it 
complementarity} relationship holds between the counts and the contribution 
to the XRB. 
If one limits the number of resolved sources in the counts by assuming, e.g., 
a stricter surface brightness selection, as discussed later on, then all 
sources pronounced unresolved will contribute to the XRB.  The joint 
consideration of these two observables is thus expected to give robust 
constraints.

\section{Results} 

In the luminosity functions $N(L,z)$ the 
luminosities are reduced to the 
ROSAT band for comparison with the data.   The derivation of  the temperature 
function $N(T,z)$ involves only eqs. (9)  
and (7), corresponding to the mere {\it passive} role played here by the ICP.  

The results depend on cosmology, and are sensitive to the values of the 
normalization $L_{o}$ and of the amplitude $\sigma_8$.  To a weaker extent, 
they depend also on the full shape of the power spectrum, 
which is determined in turn by $\Omega_0$ 
and (weakly) by $\Omega_B$, but does {\it not} depend directly on 
$\Omega_{\lambda}$ (although its normalization does).  

We note that $L_{o}$ and $\sigma_8$ decrease with $\Omega_o$ decreasing, 
 and by themselves tend to decrease all numbers; however, this is 
delicately balanced by the increase of the 
distances and by the slower (negative) 
evolution, as discussed in more detail in \S 4.4.  
With $\Omega_{\lambda}\neq 0$ the amplitude $\sigma_8$ 
 is larger, and this also enters the results as discussed in \S 4.3.

As for the cosmological parameters, we conservatively adopt combinations of 
$h$ and $\Omega_0$ yielding for the present age of the universe $t_0=13\pm 2$ 
Gyr (see Ostriker \& Steinhardt 1995).  
On the other hand, many X-ray measurements in clusters give a considerable 
ratio $\approx 0.15 \div 0.20$ of the baryons to dark matter; as rich 
clusters are 
likely to constitute fair samples of the universe, an abundance ratio 
$\Omega_B/\Omega_0=0.05\pm 0.02\,\,h^{-3/2}$ is indicated (White et al. 1993; 
White \& Fabian 1995; Markevitch et al. 1996).  
So for $\Omega_0 \approx 0.3$ a sufficient value 
$\Omega_B=0.0125\pm 0.0025\,\,h^{-2}$ is predicted by the standard cosmological 
nucleosynthesis with canonical abundances of light elements (Walker et al. 
1991); but in the critical 
case (with $h=0.5$) this must be stretched up to 
$\Omega_B=0.15\div 0.20$.  

The full $\sigma(m)$ for CDM 
cosmogonies in different cosmologies, normalized to the four-year COBE 
results (Gorsky et al. 1996), are given by Bunn \& White (1996), and 
 White et al. (1996).  
We focus on three popular CDM cosmologies/cosmogonies, which 
 provide acceptable values for $\sigma_8$: 
the tilted ($n_p=0.8$) spectrum in a critical universe with high baryonic 
content (TCDM); the scale-invariant spectrum at large scales ($n_p=1$) 
either in a flat universe 
with $\Omega_o=0.3$ and $\Omega_{\lambda}=0.7$ ($\Lambda$CDM), or in an open universe with 
$\Omega_o\approx 0.5$ (OCDM).  In the last subsection, the full set of CDM 
cosmogonies will be discussed in a more synthetic way.  

Our results will be compared  with the data from two recent surveys with 
ROSAT: the relatively local 
Brightest Cluster Sample by Ebeling et al. 1997, and the higher $z$ sample
 by Rosati et (1997). 

\subsection{Tilted CDM with high baryon content in the critical universe}

For the TCDM, we adopt the tilted primordial spectral index $n_p=0.8$ and 
 the amplitude 
$\sigma_8=0.66\,(1\pm 0.08)$, with a high baryonic fraction $\Omega_B=0.15$, 
in the critical universe with Hubble constant $h=0.5$.  

The tilt is chosen following White et al. (1996) so
 as to minimize one of the main problems 
of the Standard CDM, namely, the excess of small-scale power, still retaining 
a value for $h$ rather low but still not 
inconsistent with current observations.  In 
addition, such a cosmogony includes the high baryonic fraction referred to 
 above. 
We recall that such parameter set is hard pressed in terms of the 
low value of $h\approx 0.5$, and of the primordial abundance implied for 
light elements, with that of He exceeding many recent measurements; 
the debate is still hot on the related issue of the $D/H$ ratio, with 
recent signs of convergence, see 
Tytler, Fan \& Burles (1996); Songaila, Wampler \& Cowie (1997). 

After \S 2.4 we consider here only the PE model with the 
full range of shock strengths. 
In fig. 3a we compare the {\it local} luminosity function with 
the current data; it is seen that an acceptable fit obtains only at 
the lower $2$ standard deviations in $\sigma_8$.  
It is also seen that the evolution of $N(L,z)$ is predicted to be 
virtually nil out to $z\approx 0.8$, consistent with data by Rosati 
et al. (1997).  

The evolution of the temperature function is characterized by a fast 
decrease out to a moderate $z$ in the number of high-$T$ (massive) clusters. 
This is within
the reach of SAX (Piro et al. 1997) and  XMM (Mason et al. 1995), 
and is shown in fig. 3b. 

The counts are shown in fig. 3c.  The fit of the predicted counts to the 
bright data 
reflects the acceptability of the fit to the local 
luminosity function. Note that the slope of  predicted counts 
is sufficiently {\it flat} to fit both
the bright and the faint data by lowering $\sigma_8$ to within the COBE 
uncertainty.  Alternatively, a similar result is obtained on using the 
central value of $\sigma_8$ from COBE, but the higher baryon abundance 
$\Omega_B/\Omega_0 =0.20$.  

Our computation of the soft XRB (see fig. 3d) comprises, as said above, 
the sources fainter than $2\,10^{-14}$ erg/s cm$^2$.  Our predictions are 
compared with the residual XRB from ROSAT, once the AGNs contribution 
($\approx $ 70 \% at 1 keV, 
see Hasinger 1996) has been {\it subtracted} 
out.  The computed curve lies below the upper bounds.  

\subsection{Open CDM with $\Omega_o \approx 0.5$} 

As said, open universes with $\Omega_o\lesssim 0.3$ 
apparently constitute the simplest  way out 
of the baryonic crisis.  However, it is seen in fig. 4 
that COBE--normalized cosmogonies with $\Omega_o \leq 0.4$ using the 
adiabatic model for the ICP, suffer from the fatal flaw of severely 
underpredicting both the local 
functions of temperature and luminosity.  
The blame stays mainly with the dynamical sectors;
specifically, the low spectral amplitude $\sigma_8$ yields  
a severe deficit in $N(L,0)$, and correspondingly a deficit in the bright counts.  
Similar dynamical reasons 
 yield in aimed N-body simulations
a percentage of complex morphologies lower than 
observed in the local clusters
  (West, Jones \& Forman 1995; Mohr et al. 1995). 

On the other hand, intermediate values of $\Omega_o$ are still
a possibility.  After Liddle et al. (1996),  the range allowed 
to the class of open cosmologies by a set of observational
constraints, including $h\geq 0.5$ and the COBE normalization, is narrowed 
down to vicinity of $\Omega_o \approx 0.5$. We focus first 
on the representative OCDM cosmogony, with
$\Omega_o=0.5$, $h=0.65$ with $\Omega_B=0.07$, which yield 
$\sigma_8=0.76\,(1\pm 0.08)$.  

Here it is not easy to decide a priori whether the PE or the 
adiabatic model applies better to the ICP, so we use both, 
ending up in similar results as expected.  
These are shown in figs. 5a, 5b, 5c, 5d. It is seen that the 
local distributions are 
well fitted, but the integrated observables show excesses over the data.  

These persist when other values of $\Omega_o$ around $0.5$ are used, and when
the uncertainties in $\sigma_8$ and $L_{o}$ are considered, as discussed in 
\S 4.4.  

\subsection{CDM with $\Omega_o=0.3$ in flat geometry}

 ``Intermediate'' conditions for the cosmic deceleration also obtain 
when $\Omega_\lambda\neq 0$ is accepted, with a flat 
geometry as in most variants of inflation. The values  
$\Omega_o=0.3$ and $\Omega_{\lambda}=0.7$, with  $\Omega_B=0.05$ and $h=0.7$, 
match many observational evidences (see  Ostriker \& Steinhardt 1995). 
Following   Klypin, Primack and Holtzman 
(1997), here the normalization is $\sigma_8=1.1\,(1\pm 0.08)$. 

Here as in the previous case we have to consider both the PE and the 
adiabatic model.  We show in fig. 6a and 6b the local luminosity 
and temperature 
functions, while in figs. 6c and 6d we plot the predicted counts and the 
contribution to the soft XRB. 

\subsection {A synthetic presentation}

Here we give results covering the full set of COBE-normalized 
CDM cosmogonies. For a synthetic presentation, in figs. 7a and 7b we compare 
the predictions 
for the counts at bright ($F=2.5\,10^{-11}$ erg/s cm$^2$) 
and at faint fluxes ($F=4\,10^{-14}$ erg/s cm$^2$) with the data.  
We also show the effects of 
varying $\sigma_8$ within the uncertainty associated with COBE data, 
and $L_{o}$
within the minimum dispersion intrinsic to the $L-T$ relation. 
 Note in the figures that the strips corresponding to the uncertainties 
narrow down at 
the upper edge, because $\sigma_8$ (entering inversely the 
 expression for $\Delta N/N$) increases with $\Omega_o$. 
Similarly, $\Delta N/N$ is larger for the TCDM cosmogony 
compared with Standard CDM, 
due to the smaller value of $\sigma_8$. 

Our results agree with Liddle et al. (1996) 
in ruling out $\Omega_o<0.45$ and $\Omega_o>0.55$. 
Even in the remaining range our results compare 
critically with the 
data, because on fine--tuning $\Omega_o$ toward $\approx 
0.4$ the local luminosity function and the bright counts turn out to be 
underestimated; on the other hand, as soon as $\Omega_o\approx 0.5$ is approached, 
an acceptable fit to the local luminosity function is recovered, but an 
excess in the faint counts is generated (see fig. 5), especially with the 
adiabatic model.  In summary, agreement with {\it both} the brigth and 
the faint data is at best marginal; the 
underlying reason is that in open cosmologies long lines of sight and slow
dynamical evolution conspire to yield a slope of the counts 
too {\it steep} to account for {\it both} faint and bright counts.  

We note that such a slope would be even increased on considering that the 
formation $z$ is always larger than $z$ at observation, which has the affect of 
steepening, if only slightly, the luminosity functions (Cavaliere, 
Colafrancesco \& Menci 1993; Kitayama \& Suto 1997).  
 Note also that the addition of $\Omega_{\lambda}=0.7$ to 
$\Omega_o=0.3$ implies a
higher value of $\sigma_8$ and hence a higher levels of faint counts,
 though lower than in the $\Omega_o\approx 0.5$ case. 

In figs. 7 we also show that the counts in TCDM critical and in 
 $\Lambda$CDM cosmogonies/cosmologies
 can be made consistent with the observations on 
considering not only their uncertainties, but also those in the present COBE 
normalization and the intrinsic uncertainty in the $L$-$T$ 
relation, see the discussion following eq. (7).  

Overall, a common feature of all the above 
models based on canonical hierarchical clustering, is constituted by some 
{\it excess} in the counts; only in the critical and in the flat geometry this 
can be brought to consistency with the data.  This may indicate some 
non--trivial incompleteness in the canonical hierarchical clustering, 
worth keeping under scrutiny.  

We also show in fig. 8 
the results for the contribution to the soft XRB, e.g., at $E\approx 1$ keV.  
Once again, CDM with $0.55<\Omega_o\leq 1$ is ruled out, 
while $\Omega_o \approx 0.5$ is marginal also in this respect.  
 
Could excess faint counts be reduced by considering a stronger 
selection due to surface brightness?  On the contrary, 
we stress that the complementarity 
with the contribution to XRB makes any such excess even more significant.  
In fact, a solution cannot be sought in 
terms of surface brightness selections without increasing the excess 
contribution to the XRB from the unresolved groups and clusters. 

\section{Conclusions and discussion} 

In this paper we have computed the X-ray observables for 
groups and clusters of galaxies. As anticipated in the Introduction, we use 
-- rather than continuous and possibly 
 degenerate parametrizations --
 only discrete combinations of 
 {\it physical} models appropriate for the Dark Matter and for the 
Intra-Cluster Plasma.  

We first list our results, and then discuss them in detail. 

We have developed the punctuated equilibria (PE) model for the ICP state and 
 dynamics. This is comprised of the following two components. 

As for {\it single} clusters, we have used a polytropic $\beta$-model 
 which yields temperature profile $T(r)$ (see fig. 1) in 
good agreement 
with the observations. We predict the ICP density profile $n(r)$ and the 
 brightness profile to be flatter 
for groups than for clusters, corresponding to a 
 larger extension of the ICP relatively to their 
gravitational radii. 

As for {\it statistics}, we convolved the ICP equilibria with the histories 
of DM halos, and predicted the $L-T$ correlation to  
take the form shown in fig. 2, in agreement with the data.
In addition, we predicted an {\it intrinsic} variance 
with the minimum value also represented in fig. 2. 

Based on our PE model we then proceeded to compute for various standard 
cosmological frameworks the local and the evolved 
{\it luminosity functions} of galaxy clusters, that we compared 
with the data (fig. 3a, 4, 5a, 6a). We 
derived also the number counts (fig. 3c,  5c, 6c), 
the $z$-distributions (fig. 9) and the contribution to the soft 
X-ray background (fig. 3d, 5d, 6d). Our results are summarized 
in figs. 7a and 7b; these show that the set of acceptable 
cosmogonies/cosmologies is restricted to three disjoint {\it domains}: 
$0.4<\Omega_o<0.5$ for the standard CDM; $\Omega=1$ for the Tilted CDM; 
 $\Omega_o\approx 0.3$ for CDM in flat geometry. 
In fig. 9 we summarize the {\it confidence} levels at which 
the data are matched. 

We next proceed to discuss in detail the results listed above. 

\subsection{ICP state in evolving DM halos}

The ICP {\it state} in the hierarchically 
evolving gravitational wells constitutes the focus
of our new approach.  We propose that such state follows suit, 
 passing through a sequence of {\it punctuated equilibria} (PE) that we 
compute semi-analytically.  These computations 
comprise: the merging histories of the DM potential wells, obtained with a 
large statistics from Monte Carlo simulations of the hierarchical clustering; 
the inner {\sl hydrostatic} 
equilibrium disposition, updated after each merging episode; and the {\sl 
boundary} conditions provided by strong and weak shocks, or even by a closely 
adiabatic compression, depending on the ratio of the infall to the thermal 
energy in the preheated external medium.  

The results of our model depend on two parameters, the external temperature 
 $T_1$ and density $n_1$, which are {\sl not} free. Specifically, 
we use for $T_1$ the lower bound $T_{1\ell}=0.5$ keV 
provided by the literature on  stellar preheating; 
 in the merging events  the effective $T_1$ is the virial temperature
  of the incoming clumps, when this is larger than 0.5 keV.
  The value of $n_1$ for rich clusters 
is related to the DM density by the universal baryonic 
fraction $\Omega_B\approx 0.15$. 

Note that our PE model does not require strict spherical symmetry, but rather 
that the residual internal velocities be smaller than the inflow velocity. 
So they can include merging episodes ranging from nearly 
isotropic accretion of small clumps and diffuse gas, to anisotropic 
coalescence of comparable 
clumps along filaments of the large scale structures. 

The expression of the bolometric luminosity is proportional to 
$g^2=(n_2/n_1)^2$, the square of the density jump at the 
 bounding shock. The average of such factor over the merging histories is what 
gives to the statistical $L-T$ correlation the 
{\sl curved} shape shown in fig. 2. For rich clusters we obtain 
$L\propto T^3$. This flattens to $L\propto T^2$ for larger $T$,  
corresponding to the saturation of the shock compression factor, 
i.e., $g(T/T_1)\rightarrow 4$ when $T\gg T_1$. At the other end, the 
correlation steepens to $L\propto T^5$ in the group range, where $T/T_1\sim 1$ 
and the shocks are substantially weakened by the preheating temperature 
 in the  infalling clumps. 
The amplitude of the $L-T$ correlation  rises gently proportionally to 
$(1+z)\propto \rho_1(z)^{1/2}$ where $\rho_1$ is the density in the 
large scale structures hosting clusters and groups.   

In addition, our PE approach predicts an intrinsic {\it variance} of dynamical
origin due to the different merging histories, and built in the factor $g^2$.  
Such variance constitutes a lower bound, in view of additional 
contributions
from the variance in the ambient density, and from the central 
luminosity associated with cooling flows 
(Fabian et al. 1994; White, Jones \& Forman 1997).

\subsection{Contact with hydrodynamical simulations and with observations} 

The PE model includes, in a simplified semi-analytical form, 
compression and shocks at the boundary with the surrounding environment, 
which is modulated in density by the large scale structure  and in 
temperature by the stellar preheating. 

Simulations now clearly show the {\sl shocks} occurring also 
in major merging events 
(Schindler \& Mueller 1993; Roettiger, Stone \& Mushotzky 1997). 
The inclusion of the Rankine-Hugoniot conditions 
rises the internal temperature at the expenses of the inflow velocities. 
Complex features are found like residual kinetic pressures, 
and unmixed hot spots in the temperature distribution; but over times 
of about 2 Gyr the residual kinetic 
pressure over cluster scales reduces to less than 20 \% of the thermal one. 

On the other hand,  in the hierarchical clustering such major events are 
rare; our Monte Carlo simulations give a probability $\lesssim 20 \%$  for very 
 asymmetric events with mass ratios 2:1 or larger
 occurring within 2 Gyr from the cluster observation. 
In addition, in such events 
the ICP temperature in the infalling subcluster is comparable to 
 the virial value in the main cluster. Such major events with their low 
frequency and large $T_1$ yield 
a minor contribution to the {\sl statistical} $\langle (n_2/n_1)^2\rangle$. 

Our semi-analytical model describes only crudely 
these transient if conspicuous features, 
to focus on the lesser and more symmetric events which 
contribute the most to the $L-T$ relation.  At the 
extreme of spherical accretion of loose gas the simulations 
(see Takizawa \& Mineshige 1997)  show in detail 
that {\sl shocks} also form and expand 
slowly, to leave inside a declining temperature profile and a steeper 
density profile.  

Our PE model yields temperature profiles {\sl decreasing} 
as shown in fig. 1.  These agree 
with the published data (see Hughes et al. 1988; Honda et al. 1997; 
Markevitch et al. 1997); they agree also with   
the results from state-of-the-art simulations (Bryan \& Norman 1997) 
 obtained by running on supercomputers advanced 3D Eulerian 
codes with adapting mesh and reliable shock capturing methods. 
 While the high-resolution simulations are limited (for now and for some time 
to come) to the condition of no stellar preheating suitable only for very rich 
clusters, our model includes the effects of stellar preheating 
 of the outer gas over the whole range from groups to clusters. 

\subsection{Constraining cosmology}

With the ICP state so described, we proceeded to constrain 
the cosmological parameters. After the observations 
by Rosati et al. (1997), 
the main rule of the game turns out to be as follows: 
the dynamical evolution contained 
in the standard Press \& Schechter formula (eq. 9) must {\sl combine} with 
the evolution of the $L-T$ correlation and with cosmology to yield closely 
non-evolutionary $N(L,z)$. 

We stress that such combinations are severely selected in 
our approach.  In fact, strong {\sl shocks} are common in the {\it critical} 
cosmology, where accretion and merging activity are currently ongoing, 
and there {\sl shocks} apply in full; these include also 
weak shocks for small groups with virial temperatures below 1 keV, and for 
those rich clusters which merge with comparable clumps.  Closely 
{\sl adiabatic} compressions, instead, prevail for all structures in 
very {\it open} universes with high formation redshifts,  and little or no 
strong shocks and mixing at present.  
This defines the domain of applicability of the two adiabatic models by Kaiser 
(1991) and Evrard \& Henry (1991). 

In detail, our results are as follows (see fig. 7). 

We confirm that Standard 
CDM in the {\it critical} cosmology is 
definitely ruled out on account of its overproduction of local clusters and of
their considerable positive evolution.  
 
We also rule out {\it open} cosmologies with $\Omega_o\leq 0.3$, the 
 simplest way to enforce little evolution and also to provide
a solution to the baryonic crisis.  In fact, these cosmologies  when 
COBE-normalized 
yield a severe deficit in the local luminosity function and in the source 
counts.  

Thus, we investigated more {\it elaborate} solutions: the 
critical universe with tilted CDM and a high baryon content;  and 
two cases of intermediate deceleration, 
comprising CDM with canonical nucleosynthesis either in mildly 
open universes with $\Omega_o\approx 0.5$, or in flat geometry with 
$\Omega_o=0.3$ and $\Omega_{\lambda}=0.7$.

With the first solution we obtain (marginally) acceptable 
results (see figs. 3a-3c) for the local luminosity function $N(
L,0)$, for the source 
counts, and for the  contribution to the soft XRB, within the
present uncertainties of the data and within the variance intrinsic
to the theory.  We note that the cosmological/cosmogonical sectors by 
themselves may be tested independently,
based on the fast evolution with $z$ (see fig. 3b) 
of the temperature distribution
in the critical case, as pointed out by many authors (see Oukbir
\& Blanchard 1992; Henry 1997); this constitutes an important 
program for the satellites SAX and XMM  

As for comologies with  {\it intermediate} deceleration, 
here neither the PE nor the 
adiabatic models for the ICP are cogently indicated; thus we considered both, 
obtaining generally similar results as expected.  For open cosmologies
in particular, the results are inconsistent with the observations
of the local luminosity function and of the counts, except for the range 
$\Omega_o=0.45 \div 0.55$; even there the counts are excessive at more
than the formal $99$\% confidence level (see fig. 9), the 
excess being larger for the adiabatic models.  
The excess is due to built--in reasons, that is, the relative large amplitude 
$\sigma_8$ and the relatively steep shape of the counts, as spelled out in 
\S 4.4.  Instead, in the $\Omega_o +\Omega_\lambda= 1$ cosmology a manageable 
count excess is obtained.  

We note that the Kaiser's (1991) variant of the adiabatic models does not
yield such an excess for $\Omega_o\approx 0.5$, due to its normalization 
decreasing at high $z$.  However, the local luminosity function computed from 
this model overestimates the 
number of brightest clusters, due to the strong dependence $L\propto T^{3.5}$.  
Moreover, the normalization decrease at high $z$ is hardly consistent 
 with the data by Mushotzky and Scharf (1997). Finally, it yields 
a fast, negative evolution of $N(L,z)$ barely consistent with the 
data in the survey by Rosati et al. (1997) (the deficit is truly 
fatal in the critical or in the flat cosmology).  As luminosities 
larger than some $3\, 10^{44}$ 
erg sec$^{-1}$ are little represented in that survey, a strong test for
such a negative evolution concerns any deficit at bright fluxes in the
redshift distribution from a large--area survey.  
So we show in fig. 10 the redshift distribution  
 computed also for this model.  

We have conservatively 
chosen to focus on a limiting flux $F = 4\, 10^{-14}$ erg 
sec$^{-1}$ for which the sky coverage is nearly $100$\% (P. Rosati, 
private communication), and 
any incompleteness is out of question. Incompleteness due to 
surface brightness may be relevant at fainter fluxes, depending on the 
cluster and group profiles.  We plan to treat such issue elsewhere, but 
here we point 
out that in our approach the impact of any such incompleteness 
is limited by the {\sl complementarity} between counts of resolved 
sources and contribution to the XRB from the rest. 

Our summary is that many combinations
of standard cosmogonies/cosmologies with ICP models are ruled out.  
A relatively {\it small} set of disjoint cosmologies/cosmogonies survive, 
as shown by fig. 7:  
$\Omega=1$ with tilted CDM and high baryonic abundance combined with the 
Punctuated Equilibria, which is marginally consistent with the 
data; CDM in open cosmology with $0.5< \Omega_o<0.55$, 
which is barely consistent using the PE, and even less so using the
adiabatic models; CDM with $\Omega_o=0.3$ and $\Omega_{\lambda}=0.7$, 
which is consistent using either the PE and the adiabatic models.  

So cosmological parameters can be constrained on the basis of X--ray
clusters, but only {\it up to a point}; 
for example, the residual uncertainty in the density parameter is 
$\Delta\Omega_o/\Omega_o>20 \%$.  

\subsection{What next}

To what extent enlarging the data base on X--ray clusters will help in 
further constraining cosmology?   Here we argue that the variance 
intrinsic to the hierarchical clustering, and amplified by the ICP 
emissivity, sets an effective {\it limitation}.  In fact, fig. 7 shows that 
the present Poissonian error bars in the observed faint counts are already 
smaller than the (minimum) intrinsic variance in the predicted ones.
Decreasing the former with richer, faint surveys 
will hardly provide a sharper insight into cosmology unless one reduces both 
the uncertainty concerning 
$\sigma_8$ and the larger one concerning $L_{o}$; in fact, the two 
enter with comparable weights eq. (10), since it is 
seen that $\Delta \sigma_8/
\sigma_8$ acts roughly as $(n_{e}+3)\Delta L_{o}/6 L_{o}\approx 0.2 \Delta 
L_{o}/L_{o}$.

To what point are these reductions feasible?  On the theoretical side, the 
minimum $\Delta L_{o}/L_{o}$ of dynamical origin
may be sharpened by Monte Carlo simulations so extensive as to provide the 
full scatter  distribution.  But then one must tackle also 
the enhanced emissivity produced or signaled by cooling flows, 
correlated with higher ambient densities; this we shall treat elsewhere 
(Cavaliere Menci \& Tozzi, in preparation).  

On the observational side, one needs a large statistics for the 
distributions of $L$ and $T$; this will help in deriving narrower $L-T$ 
correlations for subsamples
categorized in term of mass deposition rates from cooling flows, see 
White, Forman \& Jones (1997).  Such aim calls for 
spectroscopic measurements of $T$, which 
are obviously harder than the bolometric $L$, and require SAX or even XMM.  
However, we stress that such efforts will find soon a more {\it proper} aim 
than constraining $\Omega_o$.  

This is because soon MAP (Bennett et al. 1997), 
and subsequently PLANCK (Bersanelli et al. 1996), will accurately measure 
on very large and still linear scales not only the perturbation power spectrum 
(from which $\sigma_8$ is derived), but also directly 
$\Omega_o$ to better that 10\%; this will supersede 
constraints set at cluster scales gone {\it non-linear}.   

Once the cosmological framework has been fixed, 
the study of groups and clusters in X-rays 
will resume what we submit to be its proper course; that is,
the physics of systems of intermediate {\it complexity}
which is comprised of the DM and of the ICP component.

With the latter fully understood and the scatter in the 
$L-T$ relation assessed, cluster X--raying will finally expose the underlying 
process of {\it non-linear} condensation of DM on scales $1-10$ Mpc. 
Then any mismatch concerning the number counts or $N(L,z)$ will be telling of 
failures either in the CDM spectra or 
in the current representation of cosmogony in terms of the 
Press \& Schechter formula. 

As a relevant example, we recall from \S 4.4 that 
even the acceptable models we computed tend to exceed the observed faint 
counts, and can be brought to consistency only at the 
lower end of the current uncertainty concerning 
$\sigma_8$. On the other hand, the 
corrections  to the Press \& Schechter formula currently discussed
yield a larger number of clusters.  For example, Jain \& Bertschinger 
(1995), and Gardner, Tozzi \& Governato (1998) find that the threshold 
$\delta_c$ must be lowered from the canonical value 1.69 to 1.5, at least at 
$z\geq 1$ if not already at $z=0$; 
a similar trend obtains considering that the formation 
redshift is always greater than the observation's as discussed in \S 4.4. 
If MAP will provide definite values of $n_p$, $\sigma_8$ and $\Omega_o$ 
such as to enhance the excess in the faint counts, then the Press \& Schechter 
rendition of the non-linear cosmogony will have to be reconsidered. 

\acknowledgements
Preliminary computations of the source counts and the contribution 
to the soft X-ray background have been performed by F. Lupini in his Thesis. 
We are indebted with P. Rosati for communicating his data prior
to publication and for many informations. We thank S. De Grandi, 
C. Gheller, M. Girardi, S. Molendi, L. Moscardini, O. Pantano and D. Trevese 
for helpful discussions.  We also thank our referee M. Henriksen
for several helpful comments, and for stimulating us to better focus
our exposition.  Partial grants are acknowledged 
from MURST and ASI.

\begin{appendix}

\section{Standard hierarchical clustering}

We recall that the variance of the perturbation field at a scale 
$R=2\pi/k$ (associated to the 
mass scale $M=\bar{\rho}\,4\pi R^3/3$ in terms of the average density 
$\bar{\rho}$) is defined by
\begin{equation}
\sigma^2(R)={1\over{V}}\int d^3\,k\,|\delta_k|^2\,W(kR)~,
\end{equation}
where $W(kR)$ is a top-hat filter (see Bond et al. 1991; Lacey \& Cole 1993).  
This is given in terms of the power spectrum $|\delta_k|^2 = k^{n_p}\,T^2(k)$ 
which depends on the physics of the early Universe (see Peebles 1993; Lucchin, 
Matarrese and Mollerach 1992) and by the subsequent microphysics. 

Results from the COBE/DMR experiment give $n_p \approx 1\pm 0.2$ 
and provide the 
normalization on large-angle scales (see Bunn \& White 1996 and references 
therein).  The transfer function $T(k)$ depends on the nature of DM 
which is the main constituent 
of the perturbations; for CDM, and given values of the
parameters $\Omega_o$, $h$ and $\Omega_B$, standard formulae are given, e.g., 
by Sugiyama (1995); White et al. (1996).  These can be recast into the form
$\sigma(m)=\sigma_8\,m^{-a}$, where $a\equiv (n_{e}+3)/6$, with $n_{e}$ being 
the effective, scale depending, index in the full power spectrum $\langle 
|\delta_k|^2\rangle$.  The values of the amplitude  at $8\,h^{-1}$ 
Mpc are given, e.g., by Bennett et al. (1996) for different cosmologies. 

The scale $8\,h^{-1}$ Mpc also defines a characteristic mass 
\begin{equation}
M_{o}={{4\pi}\over 3}\rho_o\,(8~h^{-1}{\rm 
Mpc})^3=0.6 \,10^{15}\,\Omega_o\,h^{-1}~M_{\odot}~,
\end{equation}
which we shall use as our unit mass. 

At any given mass scale, the time evolution is derived from the linear 
growth of the perturbations $D(t)$
(see Peebles 1993), which depends both on the density parameter $\Omega_o$ and 
on the parameter $\Omega_{\lambda}=\Lambda/3\,H_o^2$ associated to the 
cosmological constant $\Lambda$. An expression valid in all cases with 
$\Omega_o+\Omega_{\lambda}=1$ (as predicted by most inflationary models)
is given by Lupini (1996) in terms of the epoch $t$, and writes 
\begin{equation}
D(t)=
\Bigg[{1+3{\Omega_{\lambda}\over \Omega_o}\,\alpha^{2} \over 
{t_o^2\over t^2} + 3{\Omega_{\lambda}\over \Omega_o}\alpha^{2} }\Bigg]^{1/3}
~,
\end{equation}
where $t_o$ is the present epoch, and $\alpha\equiv 2\Omega_o^{1/2}/
\pi\,H_o\,t_o$. For $\Omega=1$ the one obtains $D(z)=(1+z)^{-1}$. 

The mass distribution $N(M,z)$ of the condensations given in eq. (9) 
has been derived 
by Press \& Schechter (1974), and is discussed by Bond et al. 
(1991). Here we stress that it contains only the linear mass variance 
$\sigma(m)$ 
and the threshold for non-linear collapse $\delta_c$ for which the canonical
value $1.69$ is taken. 

\section{Postshock conditions, and polytropic equilibrium}

The jump conditions at the shocks are based on the Rankine-Hugoniot 
conservations, and may be obtained from the implicit expressions 
given, e.g., by Landau \& Lifshitz (1959).  We work out the explicit expression
of the post-shock temperature $T_2$ for three degrees of freedom and
for a nearly static post-shock condition with $v_2<< v_1$, in the form:
\begin{equation}
kT_2={{\mu m_H v_1^2}\over 3}\Big[ {{(1+\sqrt{1+\epsilon})^2}\over 4}
+ {7\over{10}}\epsilon -{{3}\over {20}}{{\epsilon^2}\over{(1+\sqrt{1+
\epsilon})^2}}\Big]\, ,
\end{equation}
where $\epsilon\equiv 15 kT_1/4 \mu m_H v_1^2$.  In a ``cold inflow'' with 
$\epsilon <<1$ the shock is strong, and the expression simplifies to 
$kT_2\simeq {{\mu m_H v_1^2}/3} + 3kT_1/2$.  
The flow velocity $v_1$ is set 
by the potential drop across the region of nearly free fall, to read 
$v_1^2 \simeq -1.4\,  \phi_2/m_H$ where $\phi_2$ is the potential at $r=R$
(see CMT97).  Since $1.4\,  \mu\approx 1$, the above equation 
may be effectively 
approximated by $kT_2\simeq -\phi /3+3kT_1/2$\footnote{We correct here a 
numerical error in eq. (3) of CMT97, where $7/8$ appeared instead 
of $3/2$.  }.  Instead, for $\epsilon \gg 1$
the shock is weak and $T_2\simeq T_1$ is recovered as expected.  

When the temperature profile is polytropic with
$T(r)\propto n(r)^{\gamma-1}$,  eq. (2) is modified to 
\begin{equation}
L\propto T^2\,\rho^{1/2}\,
\Bigg[{T_2\over T}\Bigg]^{1/2}\,
\overline{[n(r)/n_2\big]^{2+(\gamma-1)/2}} 
\end{equation}

The ratio $n(r)/n_2$ is obtained 
starting from the hydrostatic equilibrium 
$dP/n\,dr=-d\phi/dr$ with the  polytropic pressure 
$P(r)= kT_2\,n_2\,\big[{n(r)/n_2}\big]^{\gamma}/\mu$.  This yields 
(see Cavaliere \& Fusco Femiano 1978; Sarazin 1988) the profiles 
\begin{equation}
{T(r)\over T_2}=\Big[{n(r)\over n_2}\Big]^{\gamma-1}=
1+{\gamma-1\over \gamma}\,\beta\,
\big[\tilde{\phi}_2-\tilde{\phi}(r)\big]~,
\end{equation}
where $\tilde{\phi}\equiv \phi/\mu\,m_H\,\sigma^2$ is the 
normalized potential; we use 
for $\phi(r)$ and $\sigma(r)$ the forms given
by Navarro, Frenk \& White (1996).

Eq. (B2) reduces to eq. (2) of the text in the isothermal limit 
when $\gamma\rightarrow 1$ and $T=T_2$. 
The volume--averaged factor in eq. (B2) differs from 
that for the isothermal case by 
less than $20 \%$ in the full range $1< \gamma\leq 5/3$.

\section{Luminosity functions and their integrals}

The X-ray emission of clusters of galaxies is due to optically thin, thermal 
bremsstrahlung of the hot ($T\sim 10^{7\div 8}$ K) ICP in equilibrium with the 
cluster potential wells (Cavaliere, Gursky \& Tucker 1971; see Sarazin 1988 for 
a review).  

The virial theorem provides $T\propto M/R$.  The virial radius $R$ can be 
expressed in terms of the cluster mass $M$ and the density $\rho$ to read 
$R\propto (M/\rho)^{1/3}$, which yields $T\propto M^{2/3}\,\rho^{1/3}$.  
According to the standard hierarchical clustering, the DM density inside 
clusters is proportional to the background's, so that 
$\rho \approx 200 \bar{\rho}\propto (1+z)^3$; then one obtains
\begin{equation}
T\propto M^{2/3}\,(1+z)~,
\end{equation}
corresponding to eq. (7) in the text.  The proportionality factors in eq. (C1) 
are given, e.g., by Hjorth, Oukbir \& van Kampen (1997) to imply 
$T=4.5$ keV for a cluster with $M= M_{co}$.  

The bremsstrahlung spectrum $\ell (\nu )$ at the frequency $\nu$ (in the 
frame of the source) is given by  
\begin{equation}
\ell (\nu ) \propto \int_V\,d^3r\,n^2(r)
{e^{-h\nu/kT}\over \sqrt{kT}}\,G_f(h\nu/kT)\,~,
\end{equation}
where $n$ is the ICP density, $V$ is the volume of the emission 
region and $G_f$ is the Gaunt factor, which may be effectively 
 aproximated with the function $(h\nu /kT)^{-0.4}$. From eq. (C1) and 
(C2) one obtains in terms of the observed 
frequency $\nu_o=\nu/(1+z)$ 
\begin{equation}
\ell (\nu_o )  =  {L\over{4\pi \Gamma(0.6)}}
\Big[{h\nu_o \over {kT_{co}\,m^{2/3}}}\Big]^{-0.4}\,e^{-{h\nu_o\over 
kT_{co}\,m^{2/3}}}{h\over{kT_{co}m^{2/3}}}~.  
\end{equation}
Here $L$ is the bolometric luminosity
\begin{equation}
L=L_{o}\,{\overline{[n(r)/n_1]^2}}\, m^{4/3}\,(1+z)^{7/2}~ , 
\end{equation}
where $\overline{(n(r)/n_1)^2}\equiv 
\int_0^{R}\,d^3r\,[n(r)/n_1]^2/R^3$, and $L_{o}=1.6\,10^{44}\,
h^{-2}$ erg/s corresponds to $4.5$ keV calibrated to the local $L-T$ 
correlation.  
For comparison with ROSAT data, eq. (C3) is to be integrated 
over the local range $\Delta E= 0.5 \div 2 $ keV.  The resulting 
luminosity in the band $\Delta E$ reads $L_{\Delta E}(m,z)=
w(\Delta E,m,z)\,L(m,z)$, where the correction factor is 
\begin{equation}
w(\Delta E,m,z)=\int_{E_1(1+z)}^{E_2(1+z)}\,dE e^{-E/kT(m,z)}\,G_f(E/kT(m,z))
\end{equation}

The integral number counts are given by 
\begin{equation}
N(>F)=\int dz\,dV(z)\,\int_{m_1}^{\infty} dm\,N(m,z)~,
\end{equation}
where $dV=R_H\,d\omega\,d\ell(z) D_L^2(z)/(1+z)^4$ is the cosmological volume 
subtended by the solid angle $d\omega$, $D_L(z)$ is the luminosity distance, 
and $R_H\,d\ell(z)$ is the line-of-sight element depending on $\Omega_o$ and 
$\Omega_{\lambda}$. The lower mass $m_1$ is that corresponding 
(after eq. C4 and C5) to the lowest luminosity $L_{\Delta E}$ 
detectable, at any $z$, by a survey with the limiting flux $F$. 

The complementary contribution to the soft XRB of the sources with $F' 
< F$ is given by the expression 
\begin{equation}
I(\nu_o )= \int\,dV(z)\,\int_{m_o}^{m_1} dm\;
\ell (\nu_0) {N(m,z)\over 4\pi D_L^2(z)}~.
\end{equation}
With the use of equation (C3) this yields the equation 11 of the text, 
where the limits $m_o$ and $m_1$ are discussed.  
\end{appendix}

\newpage

\figcaption[]{
Temperature (top panel) and density (bottom panel) 
profiles for a single cluster of $10^{15}\,M_{\odot}$ at $z=0$ in 
polytropic equilibrium; $\gamma=1$ (solid line), 1.1 (dotted) and 
1.2 (dashed), see eq. (B3); the dark matter potential is taken from Navarro, 
Frenk \& White (1996), and critical cosmology with TCDM perturbation spectrum 
is used.  The shock strength corresponds to 
$T_2/T_1=10$. The temperature is emission-weighted along the line of sight,
and smoothed with a filter width of 100 kpc.  
\label{fig1}}

\figcaption[]{
For the PE model and the TCDM cosmogony, we show 
the average $L-T$  relation (in terms of the emission temperatures) with its 
$2-\sigma$ dispersion (shaded region).  The average and the scatter 
of the compression factor are computed by convolving 
$g^2$ in eq. (4) with the merging histories.  The steepening 
at low $T$ is due to the  preheating temperature; this is uniformly 
distributed in the interval $0.5\pm 0.3$ keV.  
Data from Ponman et al. (1996), solid squares; David et al. 1993, 
open squares.  
\label{fig2}}

\figcaption[]{
3a: The average local luminosity function (solid line), and that 
evolved out to $z=0.7$ (dashed line) are computed for the shock model and
TCDM cosmogony.  The data are from Ebeling et al. 1997. 
\hfill\break
3b: Same as fig. 3a for the local temperature function (solid line), 
and that evolved out to $z=0.5$ (dashed line).  The data are
from Henry \& Arnaud (1991).  
\hfill\break
3c: The predicted source counts are compared with data 
by Rosati et al. (1997; solid squares) and by Piccinotti et al. (1982; solid 
triangle).  Solid line: $\sigma_8=0.66$; dotted line: $\sigma_8=0.55$.  
\hfill\break
3d: Contribution of the sources with $F< 4 10^{-14}$ erg/cm$^2$s 
to the XRB, compared  with the observed values (open stars, Hasinger et al. 
1997) with a 70\% contribution from the sources resolved by ROSAT 
subtracted out (solid squares). 
\label{fig3}}

\figcaption[]{
The local luminosity function in OCDM with $\Omega_o=0.3$ for the shock model 
(solid line) and for the adiabatic model (dashed line).  Data as in fig. 2a.  
\label{fig4}}

\figcaption[]{
Predictions for OCDM cosmogony with $\Omega_o=0.5$.  
\hfill\break
5a: luminosity functions for the shock model with the
appropriate $L$--$T$ relation: local (solid line) and evolved 
($z =0.7$, dashed line)
\hfill\break
5b: Same as 5a, for the adiabatic model.
\hfill\break
5c,5d: Same as in figs. 3c, 3d. Both the shock model (solid line) and the 
adiabatic model (long dashed line) are shown. 
\label{fig5}}

\figcaption[]{
Same as in figs. 5, but for $\Lambda$CDM cosmogony.  
\label{fig6}}

\figcaption[]{
Fig. 7a: The predicted faint counts at $F=4\,10^{-14}$ erg/s cm$^2$ are 
shown on logarithmic scales 
for the FULL range of $\Omega_o$ in Standard CDM cosmogony. The solid 
line is computed for the amplitude $\sigma_8$ of the perturbation spectrum 
corresponding to central values of the COBE normalization; the strips enclosed 
between the dashed and the dotted lines represent the uncertainties in 
$\sigma_8$ added to the intrinsic variance in $L_{o}$.  The predictions from  
TCDM cosmogony (open stars) and from $\Lambda$CDM cosmogony 
(open circle) are also shown.  The 
horizontal lines correspond to the upper and lower error bars of the data by 
Rosati et al. (1997). 
\hfill\break 
Fig. 7b: Same of fig. 7a, but for the bright counts at $F=2.5\,10^{-11}$ erg/s 
cm$^2$; here the horizontal lines correspond to the errors estimated by 
Piccinotti et al. (1982).  
\label{fig7}}

\figcaption[fig7.eps]{
The predicted contribution to the XRB (at $E\approx 1$ keV) of unresolved 
clusters and groups is shown for the whole range of $\Omega_o$ in Standard CDM 
cosmogony.  The results for TCDM and $\Lambda$CDM cosmogonies are also 
shown with same symbols as in fig. 7. 
The horizontal lines correspond to the data by Hasinger et al. (1997), with 
the 70\% contribution of resolved sources subtracted out. 
\label{fig8}}

\figcaption[fig9.eps]{
This shows the $99$\% confidence contours for both the computed local 
luminosity function (solid lines) and the computed 
number counts (dotted lines), 
in the $L_{44}-\sigma_8$ plane ($L_{44}=L_o/10^{44}$ erg/s). The 
boxes indicate $\pm 1$ standard deviations in $\sigma_8$ (corresponding 
 to the COBE uncertainty) and in $L_{44}$ (from our Monte Carlo 
$L$--$T$ relation).  Data as in figs. 3-6. 
\hfill\break 
Fig. 9a: Shock model with TCDM. This cosmogony is consistent with the counts 
within  $2$ standard deviations below $\sigma_8$. 
\hfill\break 
Fig 9b) Shock model with $\Lambda$CDM. This shows 
acceptable fits. 
\hfill\break 
Fig. 9c) Adiabatic model with OCDM  ($\Omega_o=0.5$). The 99 \% 
contours for the counts are outside the uncertainty box for $\sigma_8$ and 
$L_{44}$. 
\hfill\break 
Fig. 9d) Shock model with OCDM  ($\Omega_o=0.5$). The counts  
are consistent within $2$ standard deviations below $\sigma_8$. 
\label{fig9}}

\figcaption[fig8a.eps]{
Fig. 10a: Redshift distribution per steradian computed in the OCDM for bright 
fluxes $F\geq 10^{-13}$ erg cm$^{-2}$ sec$^{-1}$, (the corresponding
luminosity functions are given in figs. 3, 5, 6). 
Solid line: shock model; dashed line: for the adiabatic model of Evrard \& 
Henry (1991); 
dotted line: adiabatic model in the Kaiser (1991) version.  
\hfill\break 
Fig. 10b: Redshift distribution per steradian computed in $\Lambda$CDM
for fluxes $F\geq 4 \, 10^{-14}$ erg cm$^{-2}$ sec$^{-1}$.  Lines
as in fig. 10a.  
\label{fig10}}

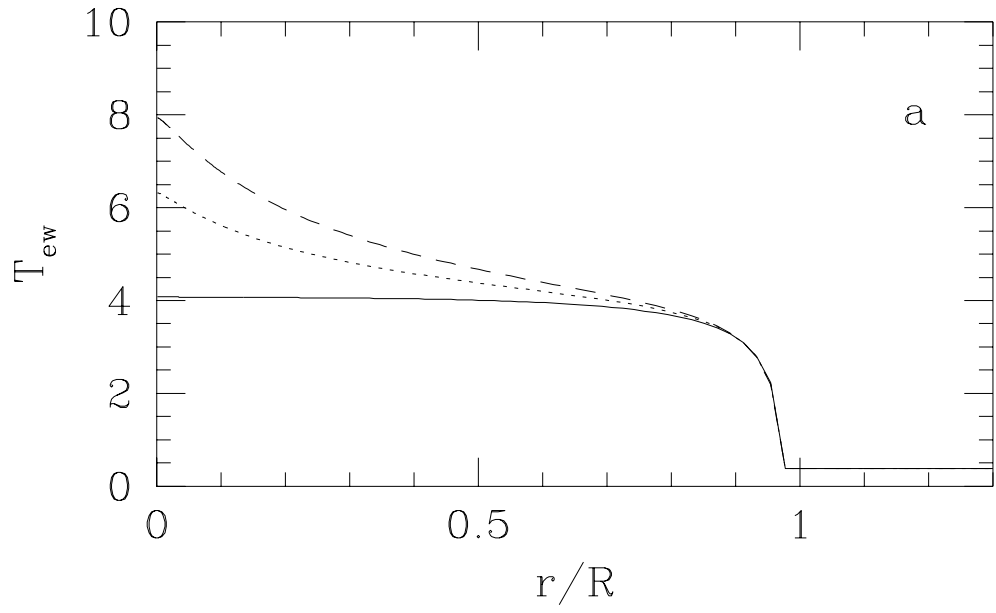
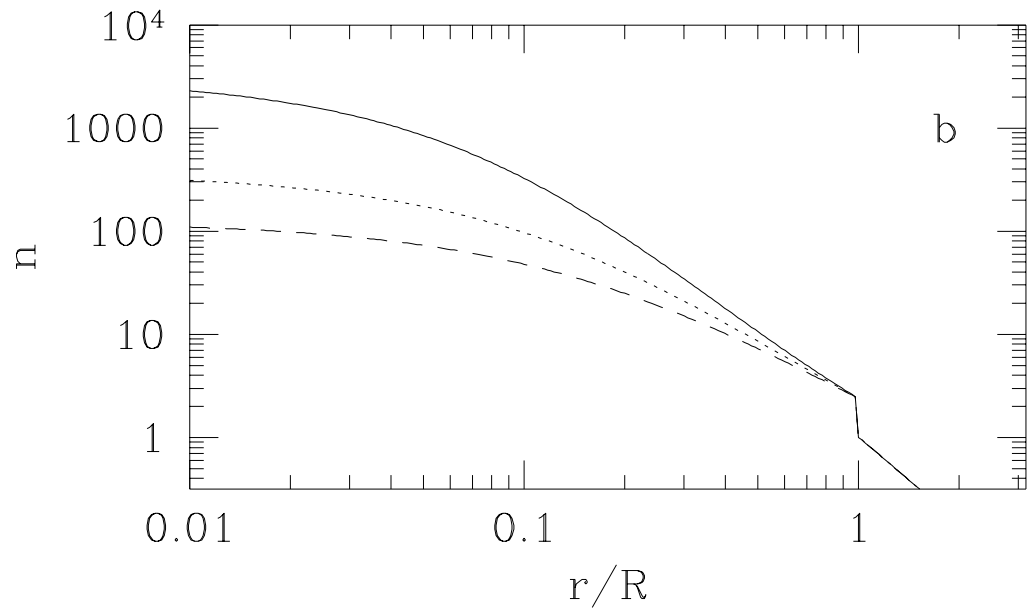
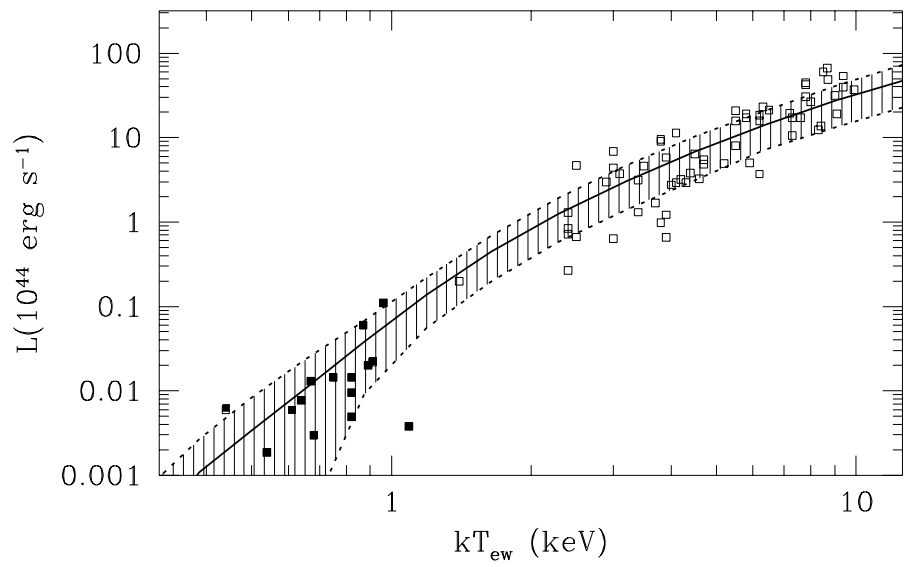
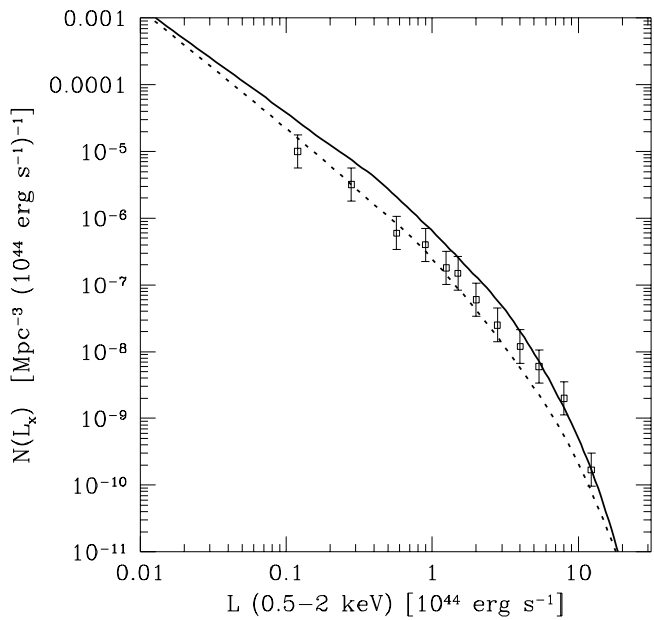
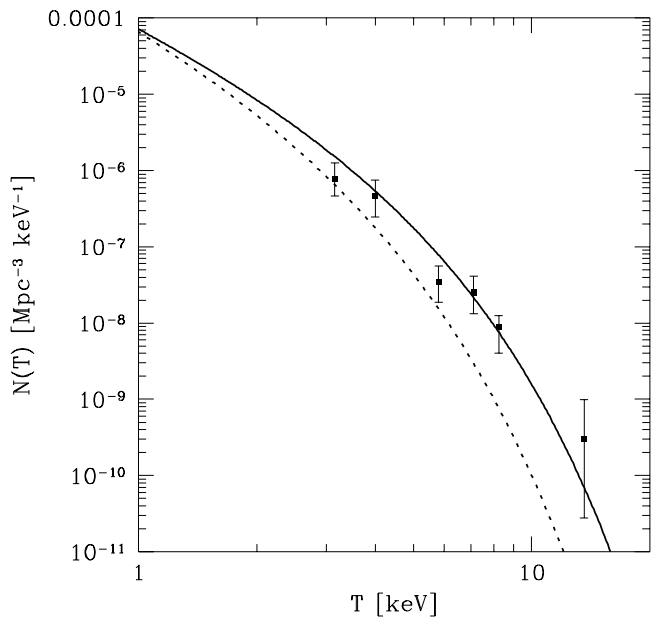
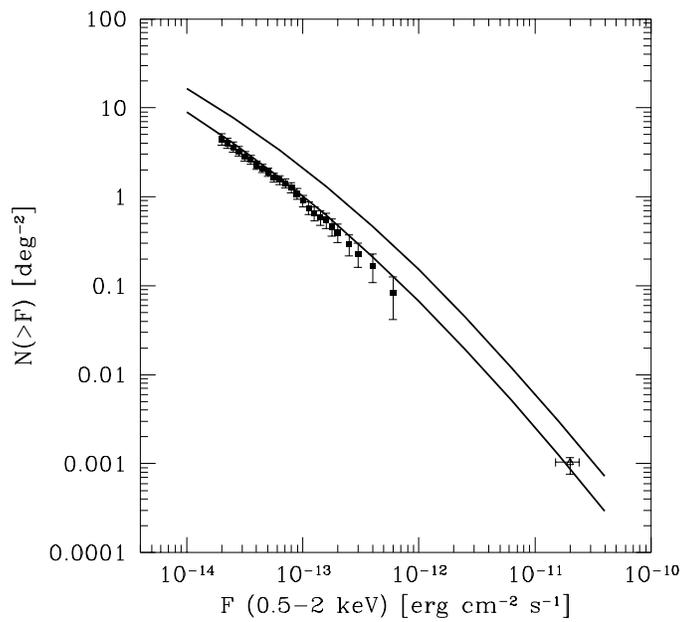
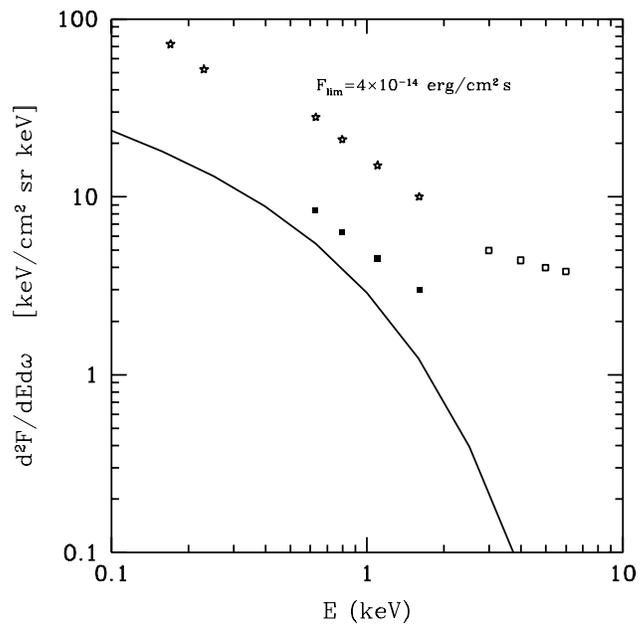
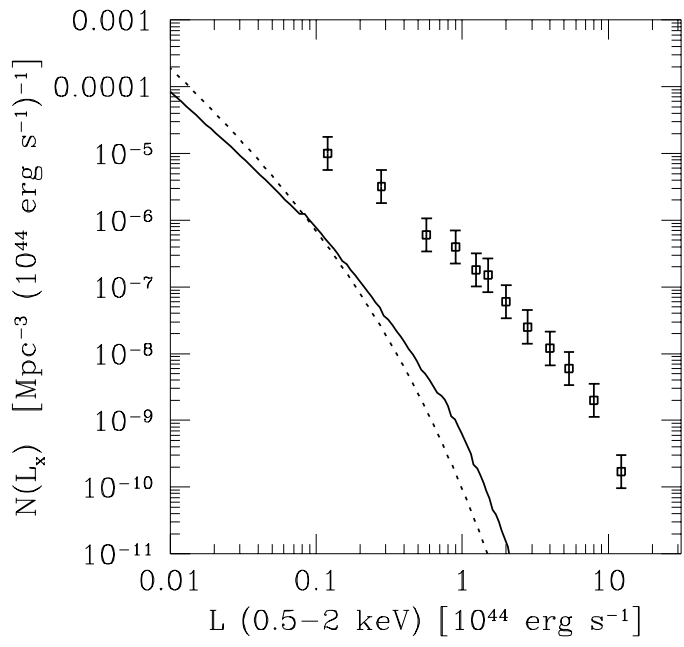
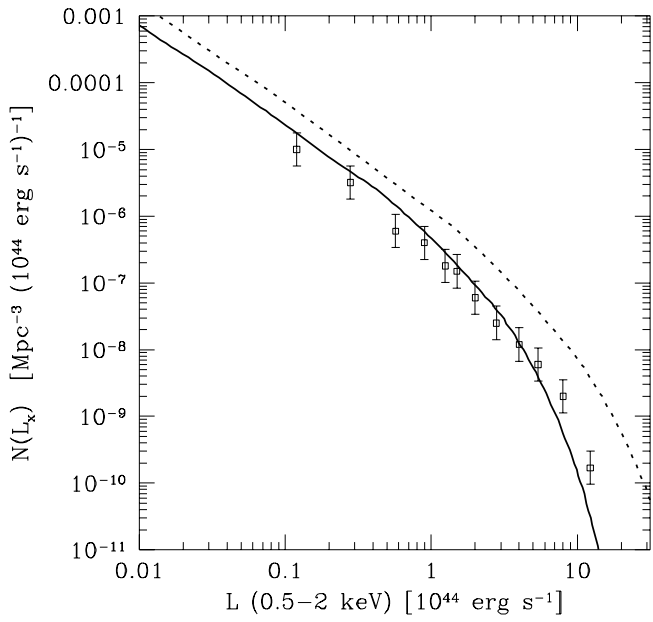
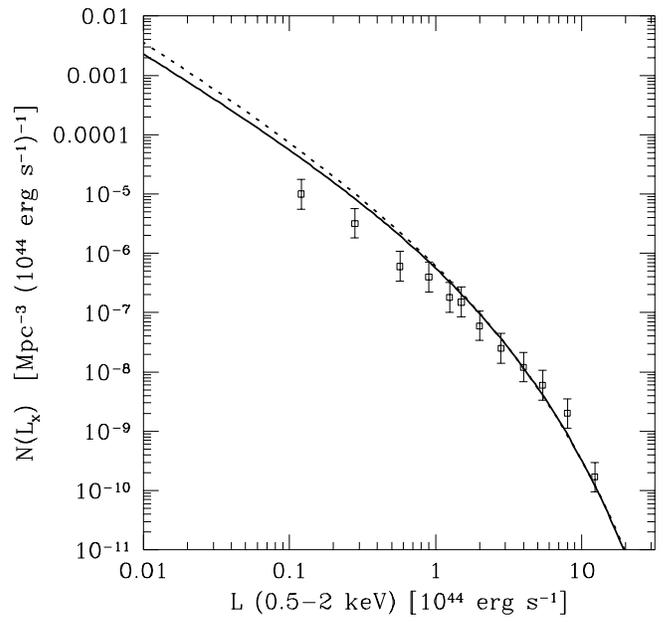
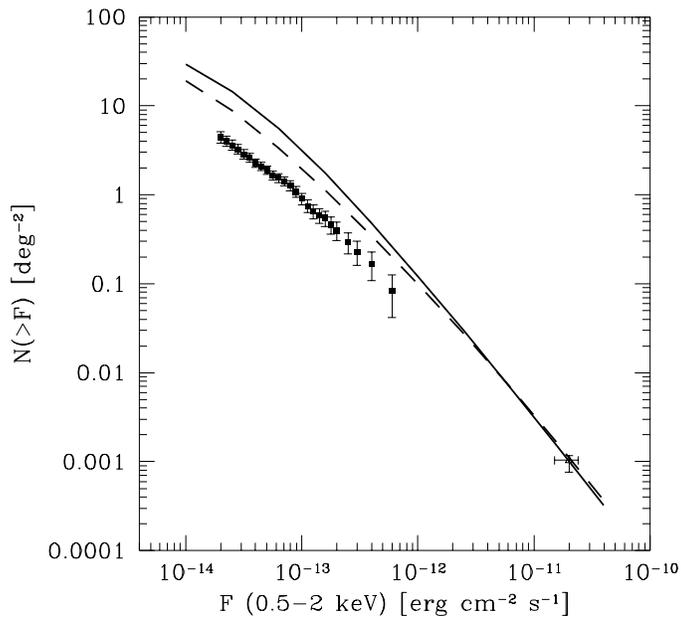
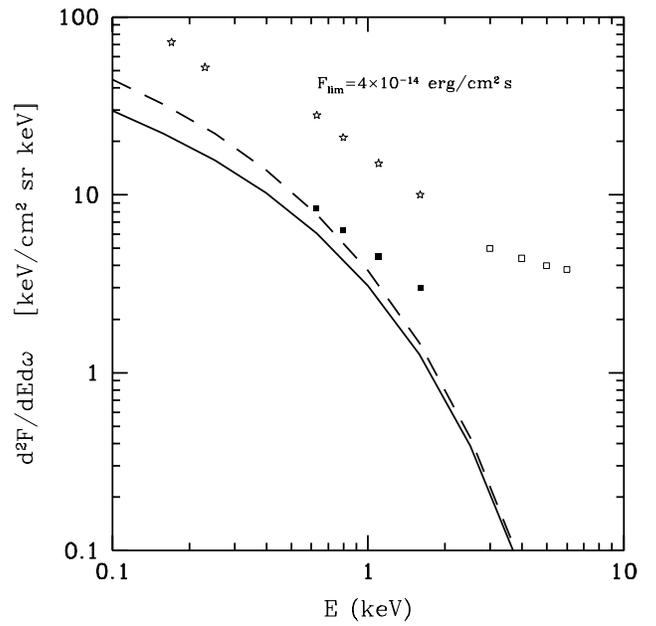
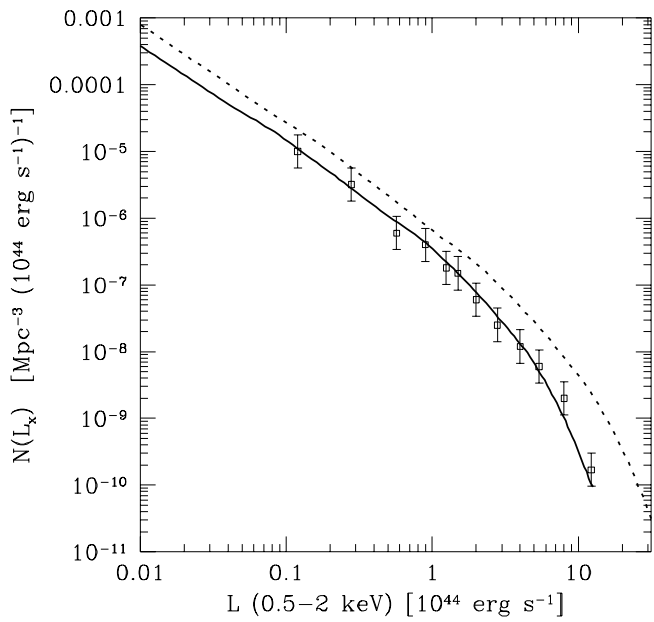
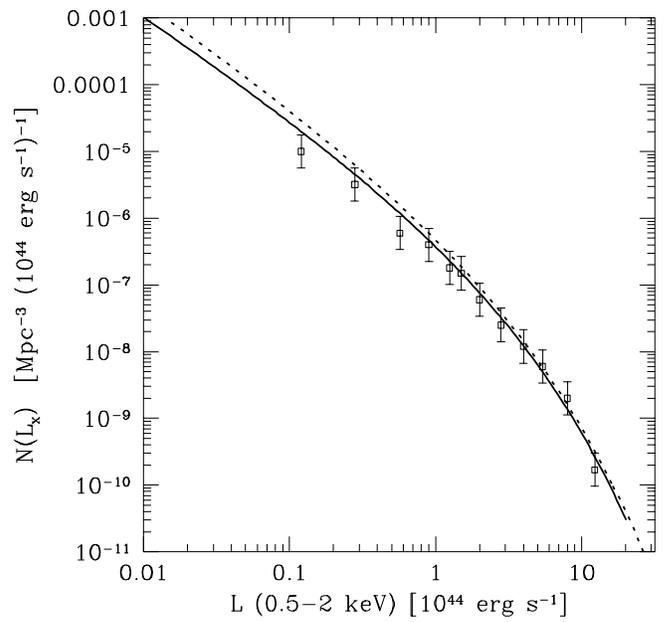
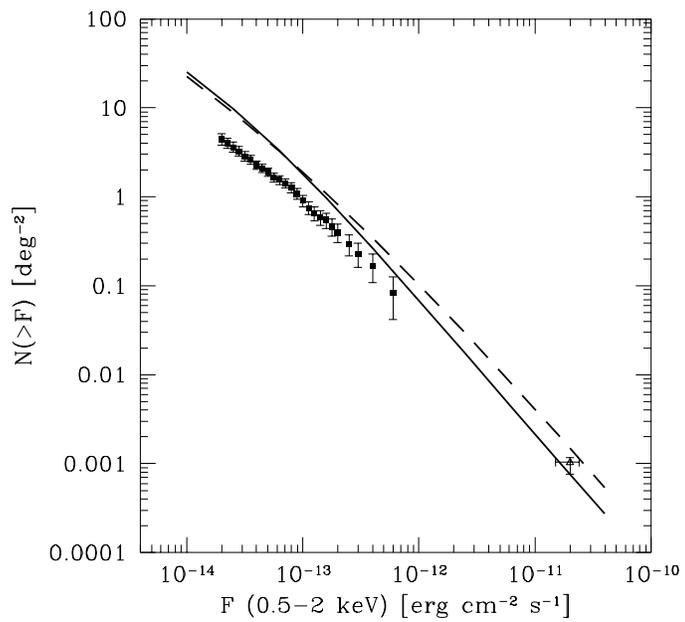
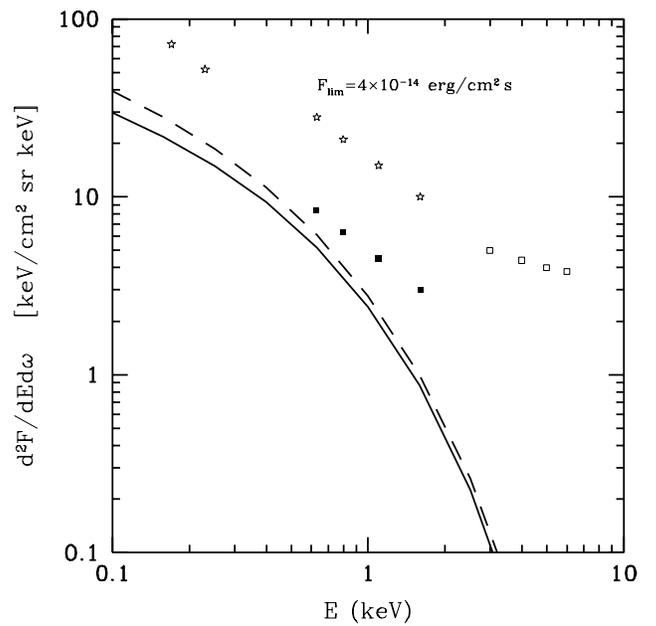
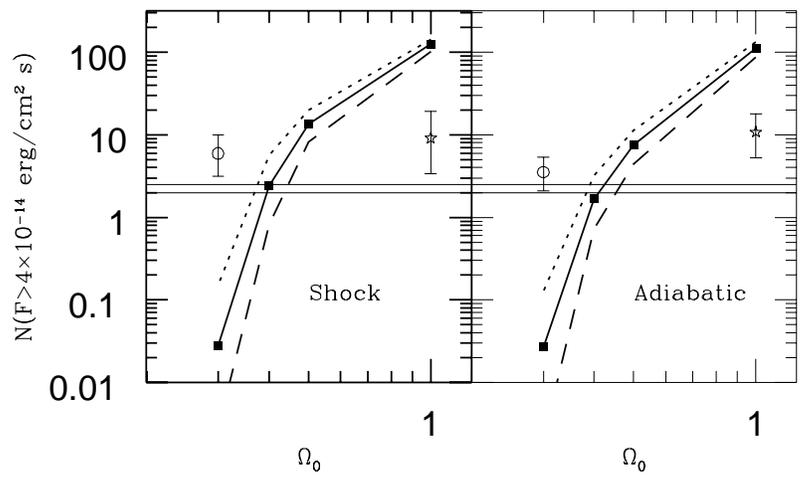
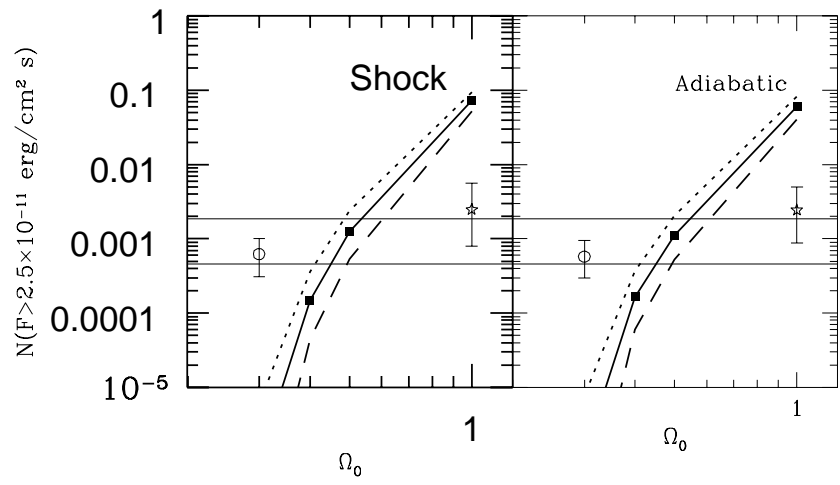
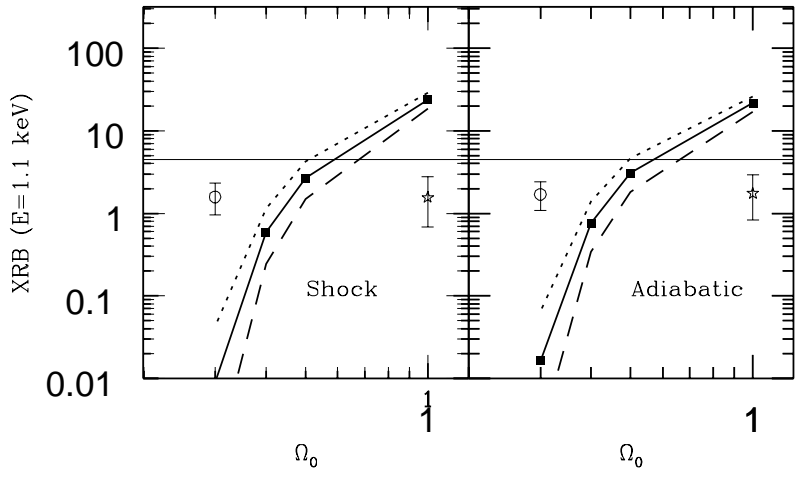
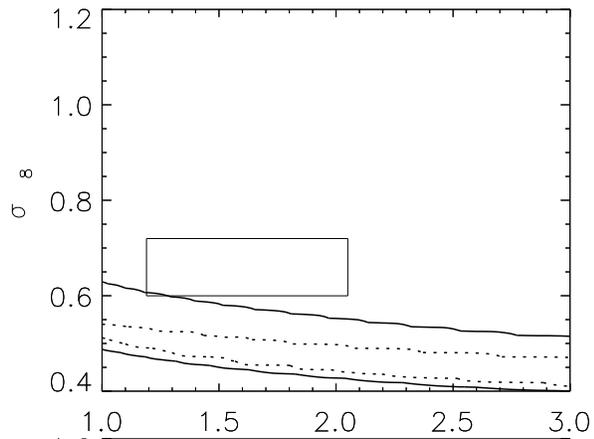
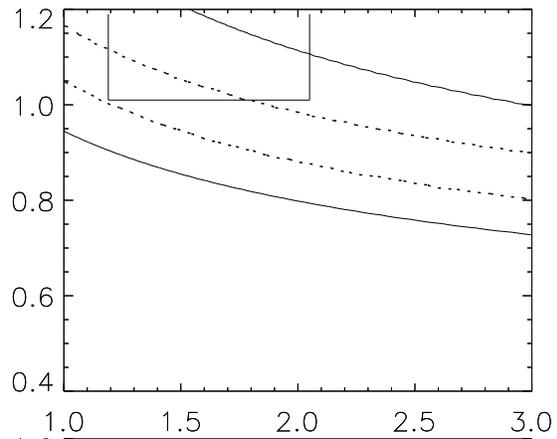
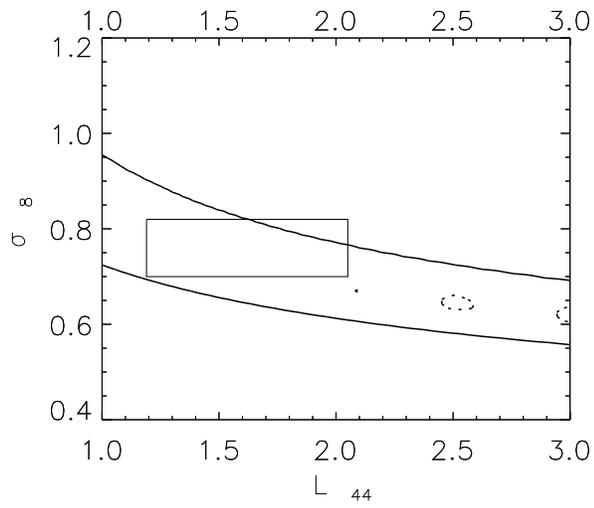
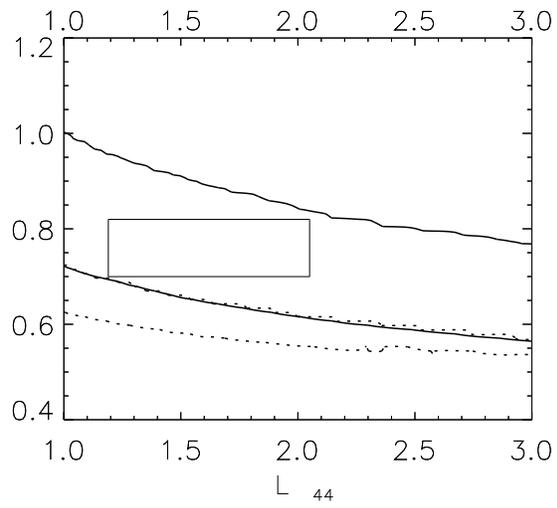
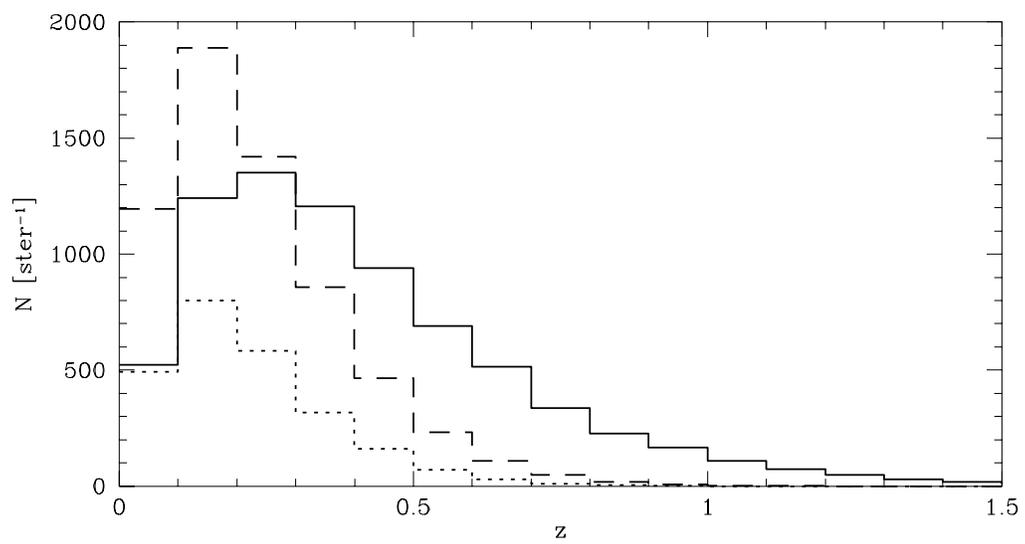
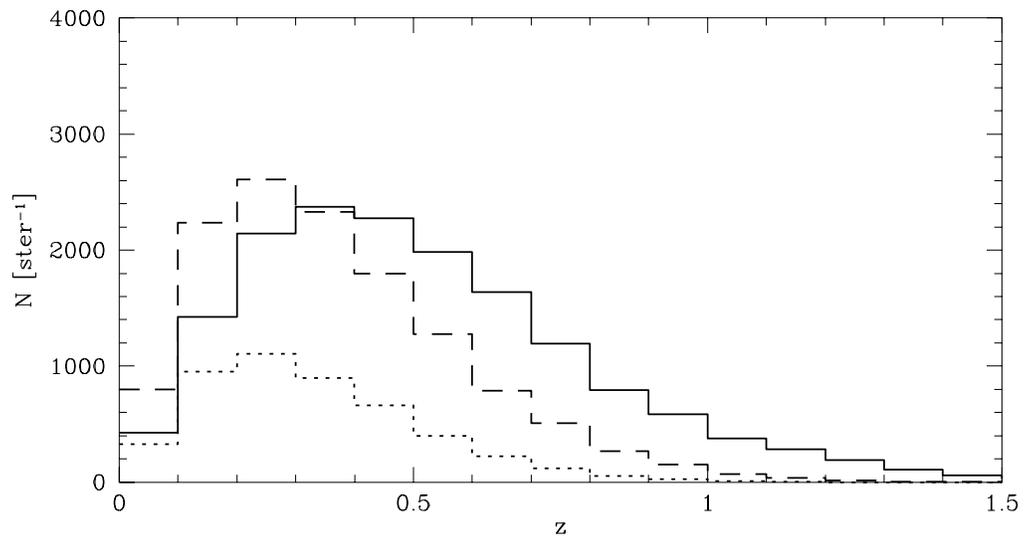
\end{document}